\documentclass[prb,preprint,superscriptaddress,nobibnotes]{revtex4}

\usepackage{graphicx}
\usepackage{dcolumn}
\usepackage{bm}
\usepackage{times,mathptmx}
\usepackage{color}
\usepackage{multirow}
\usepackage{enumitem}
\usepackage{chngpage}
\usepackage{amsmath}

\newcommand{\ave}[1]{\langle #1 \rangle}%
\newcommand{\dint} {\int\!\!\!\!\!\int}
\begin{document}


\title{\bf A silicon-chip source of bright photon-pair comb}

\author{Wei C. Jiang}
\thanks{These authors contributed equally to this work.}
\affiliation{Institute of Optics, University of Rochester, Rochester, NY 14627}
\author{Xiyuan Lu}
\thanks{These authors contributed equally to this work.}
\affiliation{Department of Physics and Astronomy, University of Rochester, Rochester, NY 14627}
\author{Jidong Zhang}
\affiliation{Department of Electrical and Computer Engineering, University of Rochester, Rochester, NY 14627}
\author{Oskar Painter}
\affiliation{Thomas J. Watson, Sr., Laboratory of Applied Physics, California Institute of Technology, Pasadena, CA 91125}
\author{Qiang Lin}
\email{qiang.lin@rochester.edu}
\affiliation{Institute of Optics, University of Rochester, Rochester, NY 14627}
\affiliation{Department of Electrical and Computer Engineering, University of Rochester, Rochester, NY 14627}



\begin{abstract}
Integrated quantum photonics relies critically on the purity, scalability, integrability, and flexibility of a photon source to support diverse quantum functionalities on a single chip. Up to date, it remains an open challenge to realize an efficient monolithic photon-pair source for on-chip application. Here we report a device on the silicon-on-insulator platform that utilizes dramatic cavity enhanced four-wave mixing in a high-Q silicon microdisk resonator. The device is able to produce high-purity photon pairs in a comb fashion, with an unprecedented spectral brightness of $6.24 \times 10^7~{\rm pair/s/mW^2/GHz}$ and photon-pair correlation with a coincidence-to-accidental ratio of $1386 \pm 278$ while pumped with a continuous-wave laser. The superior performance, together with the structural compactness and CMOS compatibility, opens up a great avenue towards quantum silicon photonics with unprecedented capability of multi-channel parallel information processing for both integrated quantum computing and long-haul quantum communication.

\end{abstract}

\maketitle

Photon-based quantum information technology has found broad applications ranging from quantum communication \cite{Gisin07}, quantum computing \cite{Kok07}, to quantum metrology \cite{Lloyd11}. Recent advances in integrated quantum photonics \cite{OBrien08, OBrien12, Osellame11, Walmsley12} show great promise for chip-scale quantum information processing with tremendous complexity. A bright, single-mode, high-purity, integrated source of single photons and/or entangled photon pairs is essential for all these applications \cite{Migdall11, Walmsley11, Pan12}, particularly for integrated quantum photonic circuits which rely critically on the purity, scalability, integrability, and flexibility of the photon source to support diverse quantum functionalities on a single chip \cite{OBrien09, Tanzilli12, Shields07}. However, although a variety of photon sources have been developed in the past few decades \cite{Migdall11, Walmsley11, Pan12, OBrien09, Tanzilli12, Shields07}, it remains an open challenge to realize an efficient monolithic source for on-chip application, lack of which current quantum photonic devices have to rely on external sources for proper operation \cite{OBrien08, Osellame11, OBrien12, Walmsley12}. Here we propose and demonstrate an ultra-bright high-purity chip-scale photon-pair source on the silicon-on-insulator (SOI) platform. By taking advantage of the dramatic cavity enhanced four-wave mixing in a high-quality silicon microdisk resonator, we are able to achieve a spectral brightness of $6.24 \times 10^7~{\rm pair/s/mW^2/GHz}$ orders of magnitude larger than other photon-pair sources \cite{Tanzilli12, Kwiat95, Wong05, Fejer07, Zeilinger07, Migdall09, Evans10, Helmy12, Eggleton12, Takesue04, Kumar05, Rarity06, Migdall07, Dyer09, Walmsley09, Sharping06, Takesue07, Baet09, Takesue10, Eggleton11, Kartik12, Thompson12, Bajoni12}, and an unprecedented photon-pair correlation with a coincidence-to-accidental ratio of $1386 \pm 278$ while pumping with a continuous-wave laser. In particular, the unique device characteristics enable photon pair generation in multiple frequency combs, thus significantly extending the wavelength management capability for integrated quantum photonics. The unprecedented device performance together with its CMOS compatibility now opens the door towards quantum silicon photonics with great potential for multi-channel parallel operation of novel quantum functionalities on chip.

To date, nearly all photon-pair sources are based upon spontaneous parametric down conversion (SPDC) or four-wave mixing (FWM) in nonlinear optical crystals/waveguides \cite{Migdall11, Walmsley11, Pan12, Tanzilli12}. Bulk crystals emit photon pairs into a multimode spatial profile, resulting in a fairly low photon generation/collection efficiency \cite{Migdall11,Tanzilli12, Kwiat95, Wong05}. Significant efforts have been devoted in recent years to developing waveguide sources for single-mode emission \cite{Tanzilli12, Fejer07, Zeilinger07, Migdall09, Evans10, Helmy12, Eggleton12}. However, the produced photon pairs generally exhibit a non-factorable spectrum which degrades considerably the quantum-state purity \cite{Walmsley11}. FWM in silica optical fibers recently appears as a promising approach with a great flexibility of engineering photon spectrum \cite{Takesue04, Kumar05, Rarity06, Migdall07, Dyer09, Walmsley09}, which, unfortunately, suffers seriously from broadband Raman scattering of silica \cite{Migdall07, Dyer09, Lin072}.

Silicon exhibits a strong Kerr nonlinearity for nonlinear optical interaction and a large refractive index enabling tight mode confinement, which has been explored intensively recently for a variety of applications \cite{Lin07, Leuthold10, Bowers10, Reed10}. In particular, single-crystalline silicon has a clean phonon spectrum with Brillouin-zone-center phonons of a well-defined frequency of 15.6~THz and a narrow linewidth of $\sim$105~GHz, which eliminates the deleterious broadband Raman noises \cite{Lin07, Lin06}. These superior features together with mature nanofabrication technology make SOI an ideal platform for integrated quantum photonic application \cite{Sharping06, Baet09, Eggleton11, Takesue07, Takesue10, Kartik12, Thompson12, Bajoni12}.

\begin{figure}[btp]
\includegraphics[width=1.0\columnwidth]{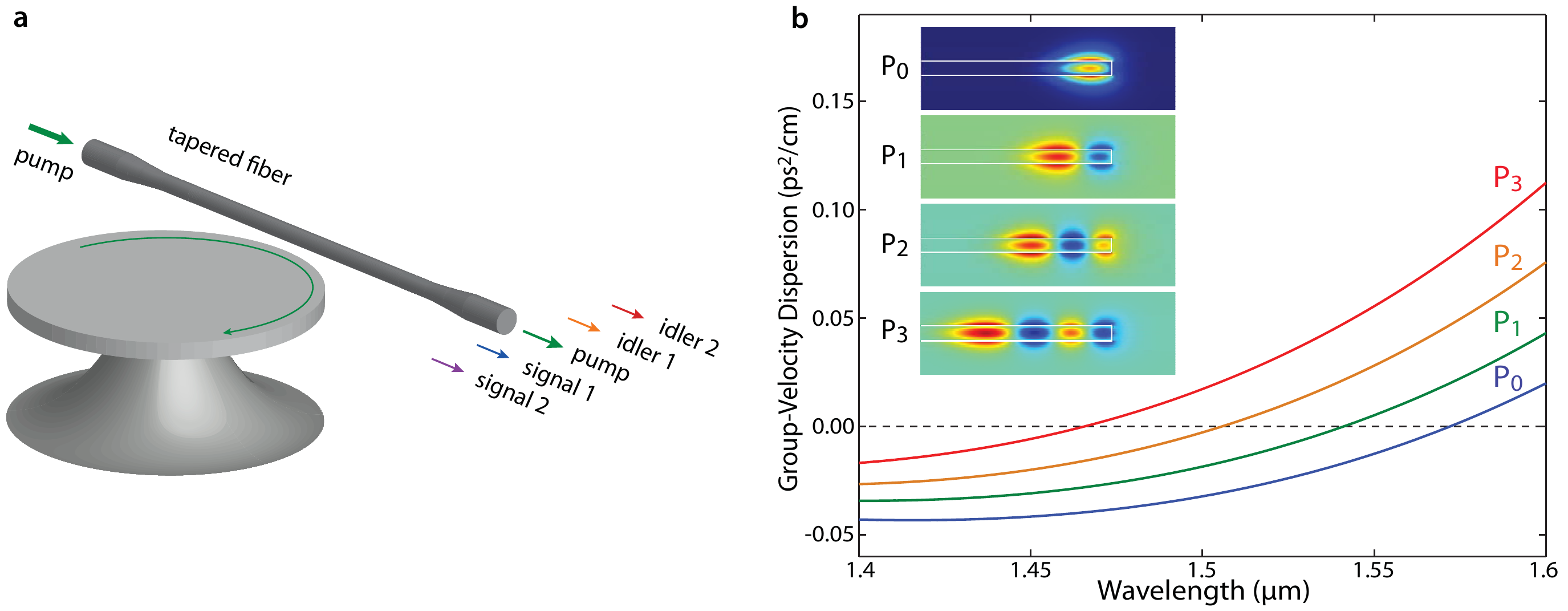}
\caption{\label{Fig1} {\footnotesize \textbf{a}, Schematic of generating a photon-pair comb from a silicon microdisk resonator sitting on a silica pedestal. All photon modes couple to the same propagation mode inside the delivery coupling waveguide, which is a single-mode tapered silica fiber. \textbf{b}, Group-velocity dispersion of vertically polarized transverse-magnetic-like (TM-like) cavity modes with four different radial orders, termed as ${\rm P_0}$, ${\rm P_1}$, ${\rm P_2}$, and ${\rm P_3}$, respectively, for a silicon microdisk with a thickness of 260~nm and a radius of 4.9~${\rm \mu m}$, simulated by the finite element method (Appendix~\ref{AppA}). The insets show the simulated optical field profiles. }}
\end{figure}

\begin{figure}[btp]
\includegraphics[width=0.720\columnwidth]{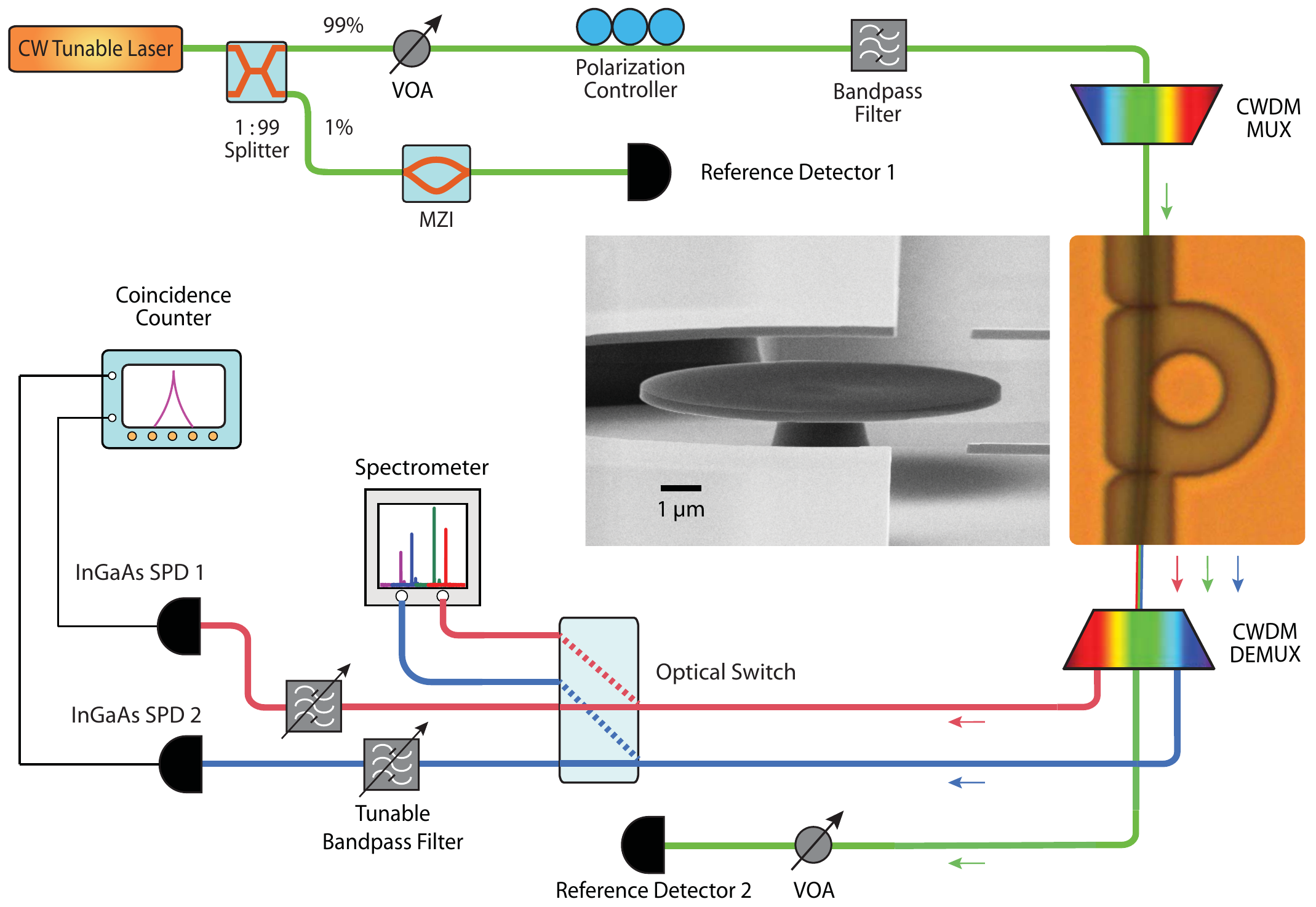}
\caption{\label{Fig2} {\footnotesize Schematic of the experimental setup for optical spectroscopy and coincidence photon counting. The pump laser is coupled into the device via a single-mode tapered silica fiber. Before coupling into the device, the pump laser passes through a bandpass filter and a coarse wavelength-division multiplexer (CWDM MUX) to cut the laser noises. The CWDM MUX has a 3-dB bandwidth of 17~nm for each of its transmission bands whose center wavelengths are separated by 20~nm apart with a band isolation of $>120$~dB. The bandpass filter is identical to one band of the CWDM MUX. The produced photon-pair comb is separated into individual photon modes by the CWDM DEMUX which is identical to the CWDM MUX used at the input end. The photoluminescence spectrum of the photon-pair comb is recorded at each transmission port of the CWDM DEMUX for easy suppression of the pump wave. For coincidence counting, the photon pairs are recorded by gated InGaAs single photon detectors (SPDs). Two tunable bandpass filters with a 3-dB bandwidth of 1.2~nm are used to cut the Raman noises produced by the delivery silica fiber. The pump power is controlled by variable optical attenuators (VOAs) and its wavelength is calibrated by a Mach-Zehnder interferometer (MZI). The inset shows the scanning electron microscopic image of the device.}}
\end{figure}

The device structure we employ for high-quality photon-pair generation is a compact silicon microdisk resonator sitting on a silica pedestal (Fig.~\ref{Fig1}a, with the scanning-electron microscopic image shown in Fig.~\ref{Fig2}). An excellent feature of a microdisk is that its group-velocity dispersion is dominantly determined by the disk thickness, allowing flexible dispersion engineering for the FWM process. As shown in Fig.~\ref{Fig1}b, with a disk thickness of 260 nm, the zero-dispersion wavelength (ZDWL) for the vertically polarized transverse-magnetic-like (TM-like) mode can be tailored to the useful telecommunication band around 1.5~${\rm \mu m}$. Detailed analysis shows that the dispersion characteristics support phase matching over a broad spectrum around 100 nm, thus enabling simultaneous generation of photon pairs in a comb fashion over all phase-matched cavity modes.

In particular, the microdisk resonator uniquely supports multiple mode families with different radial orders (Fig.~\ref{Fig1}b, inset). As they experience different mode confinement introduced by the whispering-gallery geometry, these mode families exhibit ZDWLs shifted by about 30--40~nm from each other (Fig.~\ref{Fig1}b). Consequently, a variety of desired photon-pair combs can be produced inside a single device simply by pumping at different mode families. Moreover, the complete air cladding ensures a clean interaction of the optical field with the silicon core and thus eliminates any potential contamination from the cladding material, in contrast to waveguides/microrings \cite{Sharping06, Baet09, Takesue07, Takesue10, Kartik12, Thompson12, Bajoni12} where the optical interaction with the buried oxide layer is likely to produce Raman noise photons.

Most importantly, the high-Q cavity drastically modifies the density of states of vacuum, resulting in significant Purcell enhancement on cavity modes which not only dramatically increases the generation efficiency of photon pairs but also ensures them to be created at discrete single frequencies with a high state purity \cite{Raymer05}. With $N_p$ pump photons inside the cavity, the probability of emitting a pair of signal and idler photons at time $t_s$ and $t_i$ into the coupling waveguide (Fig.~\ref{Fig1}a) is given by (Appendix~\ref{AppB})
\begin{eqnarray}
p_c(t_s, t_i) = \frac{\Gamma_{\rm es} \Gamma_{\rm ei}}{\bar{\Gamma}^2} (g N_p)^2 e^{-\Gamma_{\rm tj} |t_s-t_i|}, \label{PairProbability}
\end{eqnarray}
with $\Gamma_{\rm tj} = \Gamma_{\rm ts}$ (or $\Gamma_{\rm ti}$) on the exponent when $t_s \ge t_i$ (or $t_s < t_i$), where $\Gamma_{\rm ej}$ and $\Gamma_{\rm tj}$ (${\rm j=s,i}$) are the external coupling rate and photon decay rate of the loaded cavity, for the signal and idler photons, respectively, and $\bar{\Gamma}=(\Gamma_{\rm ts} + \Gamma_{\rm ti})/2$ represents the average. $g=\frac{c \eta n_2 \hbar \omega_p \sqrt{\omega_s \omega_i}}{n_s n_i \bar{V}}$ describes the vacuum coupling rate of the FWM process, where $n_2$, $\bar{V}$, and $\eta$ are the Kerr nonlinear coefficient, the effective mode volume, and the spatial mode overlap among the interacting photons, respectively.

As indicated by Eq.~(\ref{PairProbability}), the signal and idler photons are created in a time scale of the cavity photon lifetime within which the photon pair remains highly correlated. The longer the photon lifetime, the larger the pair emission probability. The resulting photon-pair emission rate is given by
\begin{eqnarray}
R_c = \frac{\Gamma_{\rm es} \Gamma_{\rm ei}}{\Gamma_{\rm ts} \Gamma_{\rm ti}} \frac{2 (g N_p)^2}{\bar{\Gamma}}. \label{PairRate}
\end{eqnarray}
FWM creates photon pairs inside the cavity at a rate of ${2 (g N_p)^2}/{\bar{\Gamma}}$, which are delivered to the coupling waveguide with a pair extraction efficiency of ${\Gamma_{\rm es} \Gamma_{\rm ei}}/({\Gamma_{\rm ts} \Gamma_{\rm ti}})$. A high cavity Q, a strong Kerr nonlinearity, and a small effective mode volume would result in a large photon-pair emission rate. For example, simulations by the finite-element method show that a silicon microdisk with a radius of 5~${\rm \mu m}$ exhibits a small effective mode volume of $\sim 7~{\rm \mu m^3}$ for a ${\rm P_0}$ mode, resulting in a FWM coupling rate of $g \sim 2\pi \times 460~{\rm Hz}$. Therefore, even a small pump power of $10~{\rm \mu W}$ launched into a critically coupled cavity with a Q factor of $10^6$ is able to emit photon pairs at a rate $\sim 6 \times 10^6~{\rm pair/s}$, clearly showing the extremely high efficiency of the proposed device.

The fabricated device (see Method) was tested using the experimental setup shown in Fig.~\ref{Fig2}. For easy separation of individual photon modes within the generated photon comb, the device was designed with a free spectral range (FSR) close to 20 nm such that the cavity resonances fit directly into the transmission bands of a standard coarse wavelength-division multiplexer (CWDM). The pump wave and the created photon pairs are delivered into and out of the microdisk through a tapered fiber which provides flexible control of light coupling.

\begin{figure}[btp]
\includegraphics[width=0.80\columnwidth]{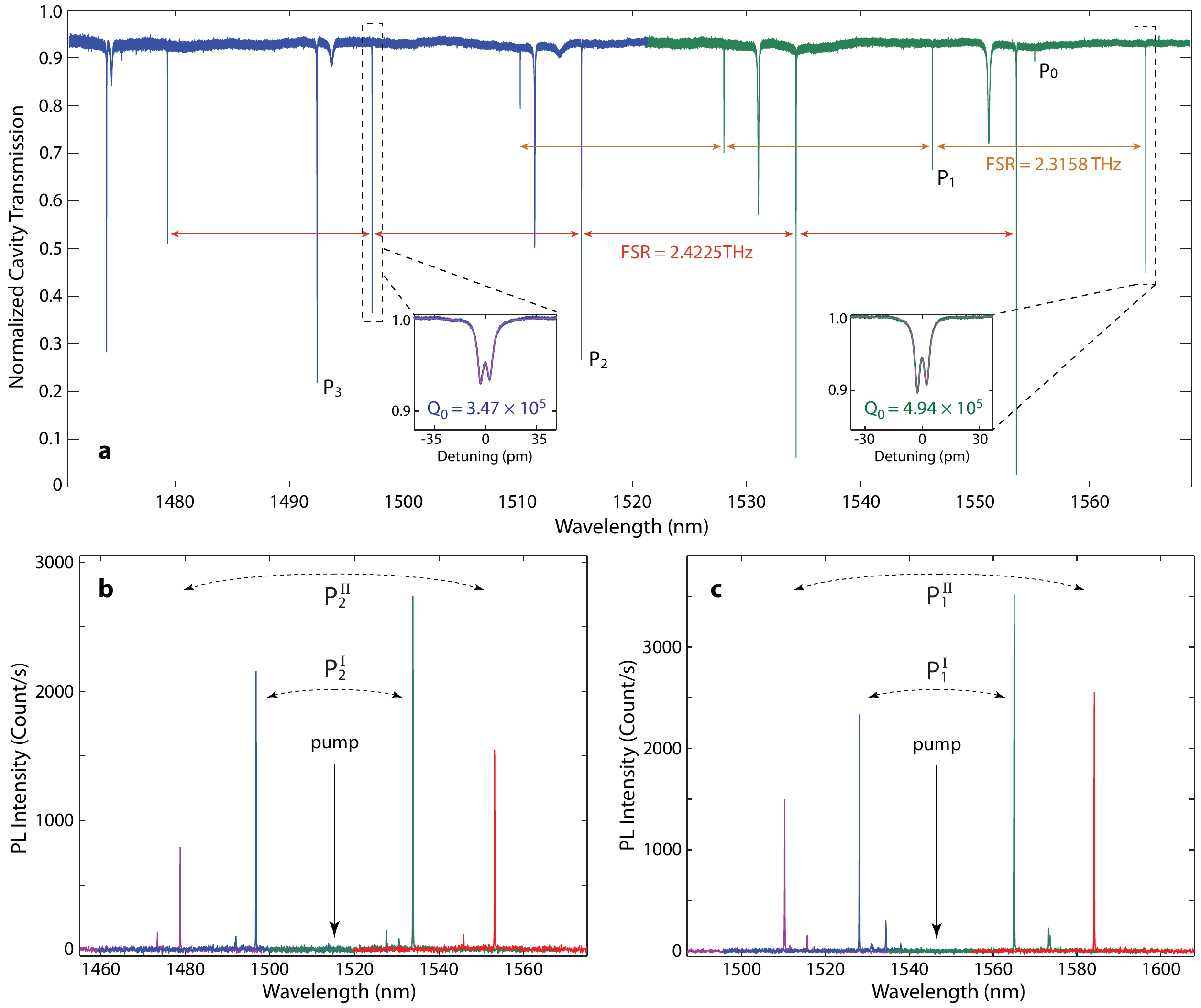}
\caption{\label{Fig3} {\footnotesize \textbf{a}, The transmission spectrum of the passive cavity scanned by two tunable lasers operating at different spectral ranges (indicated as blue and green). The two sets of arrows indicate cavity modes with constant FSRs within the ${\rm P_1}$ and ${\rm P_2}$ mode families, respectively. The insets show detailed transmission spectra of a ${\rm P_2}$ mode at 1497.2~nm and a ${\rm P_1}$ mode at 1564.9~nm, with theoretical fitting shown in red. \textbf{b,c}, The photoluminescence spectra of the ${\rm P_2}$ and ${\rm P_1}$ photon-pair comb, respectively, with the pump nearly critically coupled to the cavity. The colors indicate the individual spectra recorded at different transmission ports of the CWDM DEMUX. The four mode pairs in the two combs are termed as ${\rm P_2^{I}}$, ${\rm P_2^{II}}$, ${\rm P_1^{I}}$, and ${\rm P_1^{II}}$, respectively, as indicated on the figures. The spectrum of the ${\rm P_1}$ comb extends to the photon mode at 1584.1~nm which is beyond our laser scanning range and is thus not seen in \textbf{a}.}}
\end{figure}
The laser-scanned transmission spectrum of the passive cavity (Fig.~\ref{Fig3}a) shows four clear high-Q mode families. A constant FSR of 2.4225~THz (corresponding to $\sim$18~nm around 1.5~${\rm \mu m}$) is observed to maintain among five cavity modes for the ${\rm P_2}$ mode family, covering the spectrum from 1478 to 1555~nm. Similar phenomenon is observed for the ${\rm P_1}$ mode family with a constant FSR of 2.3158~THz, over the spectral region beyond 1510~nm. A detailed scan of the ${\rm P_2}$ mode at 1497.2~nm and the ${\rm P_1}$ mode at 1564.9~nm (Fig.~\ref{Fig3}a, insets) shows intrinsic Q factors of $3.47 \times 10^5$ and $4.94 \times 10^5$, respectively. Cavity modes within a same mode family exhibit similar optical Q since, with a same mode profile, they are of similar sensitivity to scattering loss induced by fabrication imperfection. These two mode families are used below for photon-pair generation. The ${\rm P_0}$ and ${\rm P_3}$ mode families are not used in experiment since their ZDWLs ($\sim$1572 and 1466~nm, Fig.~\ref{Fig1}b) are outside the scanning ranges of our lasers, although the ${\rm P_0}$ mode family exhibits a higher intrinsic Q factor of about $1\times 10^6$.

The constant mode spacing together with a high optical Q ensures efficient generation of a comb of photon pairs. As shown by the photoluminescence (PL) spectra in Fig.~\ref{Fig3}b, pumping at the ${\rm P_2}$ mode located at 1515.6~nm is able to produce a comb of four clean photon modes within the ${\rm P_2}$ mode family (Fig.~\ref{Fig3}b). Moreover, switching to pump at the ${\rm P_1}$ mode at 1546.3~nm is able to produce another set of clean photon comb within the ${\rm P_1}$ mode family (Fig.~\ref{Fig3}c). For the simplicity of notation, we term the photon mode pairs in the two combs as ${\rm P_1^I}$, ${\rm P_1^{II}}$, ${\rm P_2^I}$, and ${\rm P_2^{II}}$, respectively (Fig.~\ref{Fig3}b,c). The spectrum of each photon mode is so sharp that it is beyond the resolution of our spectrometer ($\sim$ 0.135~nm), implying the high coherence of generated photons. The amplitude difference within a photon comb is primarily due to different external couplings of cavity modes to the tapered fiber. When the pump mode is critically coupled to the cavity, the signals at shorter wavelengths are under coupled while the idlers at longer wavelengths are over coupled, resulting in a higher photon extraction efficiency for the idler within each mode pair (Fig.~\ref{Fig3}b,c). It also leads to a smaller overall loaded Q for the ${\rm P_2^{II}}$ (${\rm P_1^{II}}$) mode pair compared with ${\rm P_2^I}$ (${\rm P_1^I}$), thus resulting in a lower pair production rate. Note that the PL spectra also show some other modes with small amplitudes whose physical origin is not clear at this moment. Potential origins are multiphonon scattering in silicon \cite{Temple73} or light scattering from certain defects on the device surface, which require further investigation. As these tiny modes are well separated from the FWM-created photon combs, they have negligible effect on the comb performance.

\begin{figure}[btp]
\includegraphics[width=0.80\columnwidth]{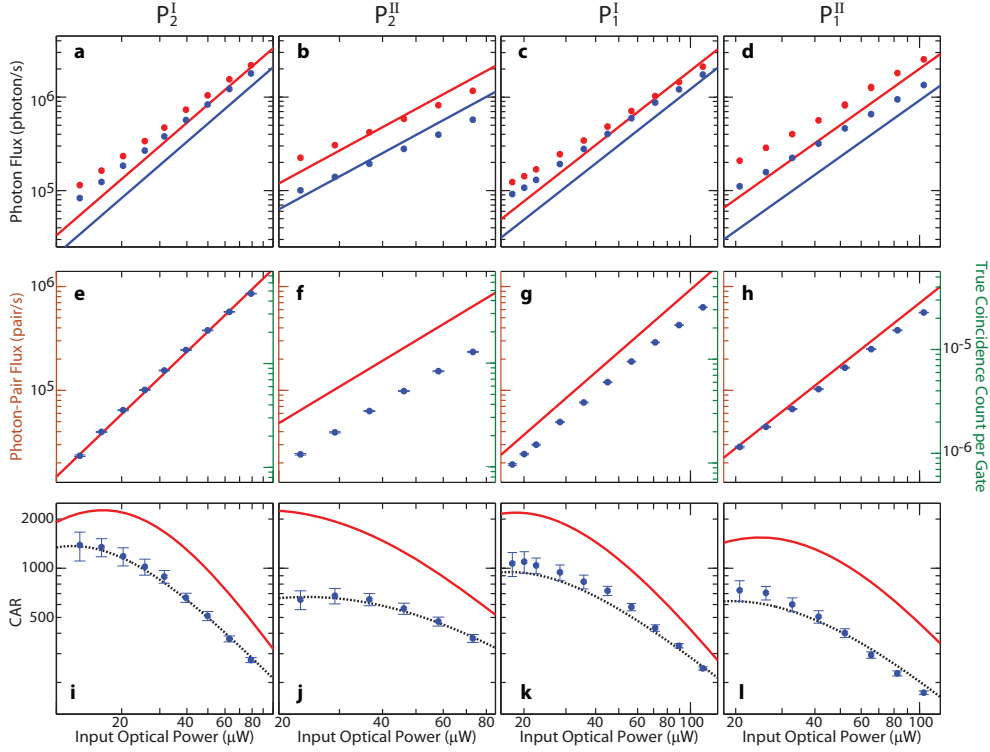}
\caption{\label{Fig4} {\footnotesize Recorded photon emission fluxes, pair emission fluxes, and CARs for the photon-pair combs. Each column of figures corresponds to one mode pair as indicated on the top. The figures within each row are plotted with identical vertical axes for each comparison among different mode pairs. \textbf{a-d}, Recorded emission fluxes of individual photon modes within each mode pair, with detector dark counts subtracted. The signal (idler) at the shorter (longer) wavelength is shown as blue (red). Solid curves show the theoretical prediction of pure FWM-created photon fluxes. \textbf{e-h}, Recorded emission fluxes of correlated photon pairs. In each figure, the left and right vertical axes show the emission flux and the corresponding true coincidence count per gate, respectively, indicated by brown and green colors. The solid curves show the theoretical prediction. \textbf{i-l}, Recorded CAR for each mode pair. No any accidental coincidence is subtracted. The red solid curves show the theoretical prediction. The black dashed curves are calculated from independently recorded true coincidence counts and photon fluxes of individual modes (Appendix~\ref{AppC}). All data were recorded with a detector gating width of 10~ns, a quantum efficiency of 15\%, a clock frequency of 250~kHz, and a dead time equal to the clock period. }}
\end{figure}
To characterize the performance of the photon-pair combs, we carried out coincidence photon counting for each mode pair when the device is pumped with a continuous-wave (CW) laser. As shown by Fig.~\ref{Fig4}e-h, the coincidence counts for all correlated mode pairs depend quadratically on the pump power, indicating they are all produced through the FWM process. The experimentally recorded pair fluxes (blue dots) are close to the theoretical predictions (red curves) (Appendix~\ref{AppB}), with a small discrepancy primarily coming from the uncertainty of Kerr nonlinear coefficient as well as measured optical Q. No significant saturation is observed, thus indicating the negligible impact of free-carrier absorption and two-photon absorption. In particular, a small input pump power of $79~{\rm \mu W}$ is able to generate a photon-pair flux of $8.55\times 10^5~{\rm pair/s}$ at ${\rm P_2^I}$ (Fig.~\ref{Fig4}e), clearly showing the high efficiency of the device. A metric characterizing the photon-pair generation efficiency is the \emph{spectral brightness}, defined as the pair flux per unit spectral width per unit pump power square (since FWM depends quadratically on the pump power). The spectral width of each photon mode is determined by the linewidth of the loaded cavity, which is 1.59 and 2.80~GHz for the ${\rm P_2^I}$ mode pair at 1497.2 and 1534.3~nm, respectively. It infers that the spectral brightness of the generated photon pairs is $6.24\times 10^7~{\rm pair/s/mW^2/GHz}$, which is about 3 orders of magnitude larger than any FWM-based photon-pair sources reported up to date \cite{Eggleton12, Takesue04, Kumar05, Rarity06, Migdall07, Dyer09, Walmsley09, Sharping06, Baet09, Eggleton11, Takesue07, Takesue10, Kartik12, Thompson12, Bajoni12}. ${\rm P_2^{II}}$, ${\rm P_1^I}$, and ${\rm P_1^{II}}$ exhibit similar spectral brightness, with magnitudes of $1.78\times 10^7$, $2.32\times 10^7$, and $2.57\times 10^7~{\rm pair/s/mW^2/GHz}$, respectively.

Indeed, the cavity enhancement is so strong inside the device that the demonstrated photon-pair source is even brighter than SPDC-based sources, although FWM, a third-order nonlinear process, is generally much weaker than SPDC which is a second-order nonlinear process. The spectral brightness of SPDC-based sources is generally described by the slope efficiency given as the photon-pair flux per unit spectral width per unit power \cite{Wong05, Fejer07, Zeilinger07, Migdall09}. ${\rm P_2^I}$ shown in Fig.~\ref{Fig4}e has a slope efficiency of $4.93 \times 10^6~{\rm pair/s/mW/GHz}$ at a pair flux of $8.55\times 10^5~{\rm pair/s}$, which is more than one order of magnitude larger than any SPDC-based waveguide sources \cite{Tanzilli12, Kwiat95, Wong05, Fejer07, Zeilinger07, Migdall09, Evans10, Helmy12}. It is comparable with the state-of-the-art cavity-enhanced SPDC sources \cite{Benson09, Pan11, Harris12}, while the latter create photon pairs in multi-spatiotemporal modes which require stringent filtering. Moreover, the latter are constructed with bulk optical components which require sophisticated cavity locking and are challenging for chip-scale integration.

Figure \ref{Fig4} shows that the pair fluxes are smaller than those of corresponding individual photon modes. This is because of the non-unity photon extraction efficiency, a common feature of cavity quantum electrodynamic systems: Although the signal and idler photons are always created in a pair fashion inside the cavity, they transmit out of the cavity independently, only a certain fraction of which remains the pair correlation. The performance can be improved by using cavity modes with higher intrinsic Q (\emph{e.g.}, the ${\rm P_0}$ mode family) together with higher external coupling. On the other hand, Fig.~\ref{Fig4}a-d show that the fluxes of individual photon modes deviate slightly away from the quadratic power dependence, indicating that they are accompanied with a certain amount of noises. Detailed characterizations show that the noise photons dominantly come from the broadband Raman noises generated inside the delivery fibers (Appendix~\ref{AppD}). In experiment, a pair of bandpass filters with a 3-dB bandwidth of 1.2~nm were used in front of the single photon detectors (SPDs) (Fig.~\ref{Fig2}) to suppress the Raman noises. As each photon mode has a very narrow spectral width, the Raman noises can be suppressed further by using a narrower filter. They can also be reduced by shortening the lengths of delivery fibers. Such Raman noises would be absent if an on-chip laser \cite{Bowers10} is used to generate the photon-pair combs.

An important figure of merit characterizing the quantum correlation between the photon pairs is the \emph{coincidence-to-accidental ratio (CAR)} \cite{Takesue10, Migdall09, Rarity06, Dyer09}, which is directly related to the quantum-interference visibility of the entangled states constructed from the photon pairs \cite{Lin072}. Figure~\ref{Fig4}i-l show the recorded CAR (blue dots) for all mode pairs. In general, CAR increases with decreased photon-pair flux because of reduced probability of multi-pair creation and saturates at a certain level of pair flux when the detector dark counts start to dominate. All mode pairs exhibit very high CAR at all power levels. For the ${\rm P_2^I}$ mode pair, the device is able to achieve unprecedented CAR with a value of $274\pm10$ at a pair flux of $8.55\times 10^5~{\rm pair/s}$ and a peak value of $1386 \pm 278$ at a pair flux of $2.35 \times 10^4~{\rm pair/s}$. These values are more than three times larger than other photon-pair sources \cite{Zeilinger07, Migdall09, Evans10, Helmy12, Eggleton12, Takesue04, Kumar05, Rarity06, Migdall07, Dyer09, Walmsley09, Sharping06, Baet09, Eggleton11, Takesue07, Takesue10, Kartik12, Thompson12, Bajoni12}, even without using pulsed pump to suppress noise photons \cite{Migdall09, Rarity06, Migdall07, Sharping06, Eggleton11, Takesue07, Takesue10} or delicate superconducting SPDs to reduce detector dark counts \cite{Fejer07, Dyer09, Thompson12, Bajoni12}. Similar CAR is achieved for ${\rm P_1^I}$, with values of $1097 \pm 165$ and $244 \pm 6$ at pair fluxes of $2.43 \times 10^4$ and $6.31 \times 10^5~{\rm pair/s}$, respectively. At the same time, ${\rm P_1^{II}}$ and ${\rm P_2^{II}}$ exhibit as well relatively high CAR, with peak values of $733 \pm 106$ and $679 \pm 74$, respectively. For all the mode pairs, the power-dependent trend of CAR follows closely the theory (red curves) (Appendix~\ref{AppC}). The discrepancy primarily comes from the uncertainty in characterizing Raman noises generated in delivery fibers. These results clearly show the superior quality of generated photon-pair combs. We expect that the peak CAR of the photon-pair combs can be improved by at least one order of magnitude in the future, by suppressing the Raman noises produced in the delivery fibers and by using better detectors (\emph{e.g.}, superconducting SPDs) with lower dark counts.

\begin{figure}[btp]
\includegraphics[width=0.67\columnwidth]{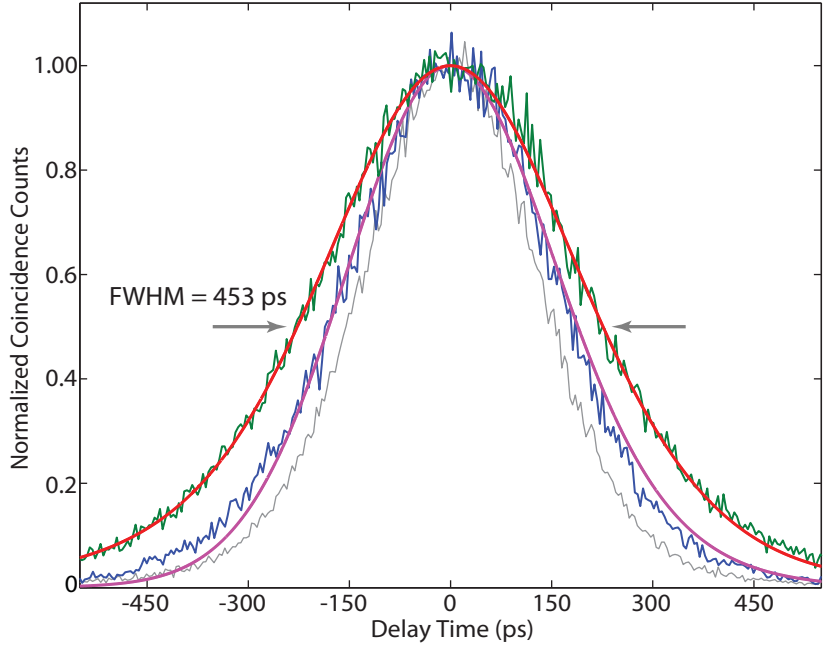}
\caption{\label{Fig5} {\footnotesize Normalized delay-time-dependent coincidence spectra for ${\rm P_2^I}$. The blue and green curves show the cases with different external coupling conditions and thus different averaged photon lifetimes. The gray curve shows the instrument response function (IRF) of our coincidence counting system, with a FWHM of 312~ps which primarily come from the timing jitters of the two SPDs. The red and purple curves show the theoretical prediction. The coincidence spectra were recorded with a detector gating width of 10~ns, a quantum efficiency of 25\%, and a clock frequency of 125~kHz. The coincidence counter has a time resolution of 4~ps. }}
\end{figure}

The high pair correlation and high coherence of the photon-pair combs allow resolving the temporal structure of quantum correlation between the signal and idler photons. This is particularly enabled by the tapered-fiber coupling system which provides flexible control of the external light coupling and thus the cavity photon lifetime. Figure \ref{Fig5} shows the normalized coincidence spectra of ${\rm P_2^I}$ with two different external coupling conditions. When the tapered fiber is close to the device, the external coupling is strong which results in a short photon lifetime of 65~ps (averaged between signal and idler). As a result, the coincidence spectrum (blue curve) shows a full width at half maximum (FWHM) of 370~ps, which is 19\% larger than the time resolution of the instrument response function (IRF, gray curve) of our coincidence counting system. However, moving the tapered fiber away from the device increases the photon lifetime considerably to 171~ps, leading to a coincidence spectrum (green curve) with a FWHM of 453~ps which is 45\% larger than the time resolution of IRF. The experimentally observed time correlation is explained well by the theory (Appendix~\ref{AppC}). To the best of our knowledge, this is the first time to resolve the temporal photon correlation for a chip-scale photon-pair source, although it was observed before in cavity-enhanced SPDC sources \cite{Benson09, Pan11, Harris12}. Quantum coherence between correlated photon pairs is essential for constructing a variety of quantum functionalities \cite{Gisin07, Kok07,Pan12}. The photon lifetime of 171~ps in our device corresponds to a propagation length of $\sim1.5$~cm, which is about the footprint of a photonic chip. Therefore, the produced photon pairs are able to share temporal coherence over the entire chip, which would significantly improve the quality of quantum functionalities built upon.

The superior performance of the demonstrated device offers a great opportunity for realizing a variety of high-quality quantum functionalities on chip. Straightforward applications include producing heralded single photons \cite{Migdall11} and quantum entanglement in various degrees of freedom (frequency \cite{Zeilinger09}, path \cite{Pan12}, time \cite{Gisin07}, polarization \cite{Takesue10}, etc.), which are essential for quantum communication and quantum computing \cite{Gisin07, Kok07}. Compared with conventional approaches, the high photon coherence in the device would significantly improve the purity of the constructed quantum states. As the cavity enhancement only occurs within the cavity photon lifetime, the demonstrated approach is able to operate with a repetition rate above gigahertz by use of pulsed pumping. Further increase of photon-pair coherence and spectral brightness can be obtained by improving optical Q (\emph{e.g.}, through surface passivation \cite{Painter06}). Such a high coherence would enable teleporting quantum entanglement intra-chip, inter-chip, or off-chip over far distance free from the stringent synchronization requirement in conventional schemes \cite{Gisin07, Pan12, Gisin072} (The direct photon coupling into optical fibers with negligible losses (Fig.~\ref{Fig2}) is particularly suitable for this application). Moreover, the comb nature of emitted photons makes readily available wavelength-division multiplexing in quantum regime, and thus significantly extends the spectral efficiency and wavelength management capability of integrated quantum photonics. In particular, it exhibits great potential for realizing multipartite entanglement \cite{Pan12} and cluster state \cite{Pfister11} inside a single device, which would dramatically enhance the capability of integrated quantum computing. On the other hand, the high coherence and spectral flexibility of the demonstrated approach enable excellent interfacing with other chip-scale information processing/storage elements \cite{OBrien09}. These features together with the developed silicon photonic technology \cite{Bowers10, Reed10} offer a great opportunity to ultimately form a complete CMOS compatible chip-scale platform for quantum information processing.

\vspace{-0.6cm}
\section*{Methods}
\vspace{-0.5cm}
\begin{footnotesize}
The silicon microdisk resonator is fabricated from a standard silicon-on-insulator wafer by use of e-beam lithography to define the device pattern, fluorine-based plasma to etch the silicon layer, and hydrofluoric acid to undercut the silica pedestal. Light is coupled into and out of the microdisk through a tapered optical fiber which is anchored onto two nanoforks fabricated near the microdisk for stable operation (see Fig.~\ref{Fig2}). Touching to the nanoforks introduces a small insertion loss of only about 7\%, as indicated by the cavity transmission spectrum given in Fig.~\ref{Fig3}a. The PL spectra were recorded at the transmission ports of the CWDM DEMUX to suppress the residual pump wave which otherwise would cause significant stray light inside the spectrometer.

The CAR is obtained by the following procedure: First, at each pump power, a delay-time-dependent coincidence counting histogram (we term as coincidence spectrum here for simplicity. See Fig.~\ref{Fig5} for an example) is obtained with zero time delay between the two SPDs. The coincidence spectrum is integrated over its full width at half maximum $\tau_{FWHM}$ to obtain the total coincidence $C$. The statistical error of $C$ is given by $E_C = \sqrt{C}$ since the photon counting follows a Poisson distribution. To find the accidental coincidence, a series of coincidence spectra are obtained with a time delay of $j T_0$ between the two SPDs, where $T_0$ is the detector clock period and $j=1,2,...,200$. Integrate each of them within the same $\tau_{FWHM}$ to obtain an accidental coincidence $A_j$. The average and the standard deviation of $A_j$ give the accidental coincidence $A$ and its statistical error $E_A$. Note that no any accidental coincidence is subtracted. Finally, the CAR is given by ${\rm CAR} = (C-A)/A$ and its statistical error $E_{\rm CAR}$ is given by $\frac{E_{\rm CAR}}{\rm CAR} = \sqrt{\left(\frac{E_C}{C}\right)^2 + \left(\frac{E_A}{A}\right)^2 }$. To reduce the statistical error, each data point in Fig.~\ref{Fig4} was recorded with a time period of $5400-18900$ seconds depending on the magnitude of photon-pair flux. With such amount of photon counting data, $\frac{E_C}{C}$ is very small ($<2$\%) and $\frac{E_{\rm CAR}}{\rm CAR}$ is dominantly determined by $\frac{E_A}{A}$ since the accidental coincidence is very small in our device.

To find the photon-pair emission flux shown in Fig.~\ref{Fig4}e-h, a same procedure is carried out but now the integration is performed over the entire time window of coincidence spectra (20~ns, with a detector gating width of 10 ns) to obtain the total coincidence $C_t$, the accidental coincidence $A_t$, and their statistical errors $E_{C_t}$ and $E_{A_t}$. The true coincidence is thus given by $C_T = C_t - A_t$. Its statistical error $E_{C_T}$ is given by $E_{C_T} = \sqrt{\left(E_{C_t}\right)^2 + \left(E_{A_t}\right)^2 }$. Dividing $C_T$ by the total number of detector gates within the data collection time we find the true coincidence counts per gate (which is shown on the right axes of Fig.~\ref{Fig4}e-h). The photon-pair emission flux is obtained by calibrating the true coincidence counts per gate with the clock frequency, duty cycle, and quantum efficiency of the detectors as well as the propagation losses from the device to the detectors for the photon pairs. The statistical error of pair emission flux is obtained accordingly from $E_{C_T}$.

The IRF of our coincidence counting system is obtained with identical ultrashort pulses launched onto the two SPDs. The ultrashort pulses come from a mode-locked fiber laser with a pulse width (FWHM) of 240~fs. They are broadened to about 5~ps by the delivery fibers before they hit on the SPDs. The pulse width is much shorter than the timing jitters of the SPDs, thus providing a fairly accurate approach to characterize the IRF.

In all photon counting measurements, the detector gating width is set to 10 ns and the deadtime is equal to the clock period. The two InGaAs SPDs have dark counts of $5.09 \times 10^{-5}$ and $3.18 \times 10^{-5}$ per gate, respectively, at a quantum efficiency of 15\% and a clock frequency of 250 kHz. The detector dark counts increase to $8.96 \times 10^{-5}$ and $6.93 \times 10^{-5}$ per gate at a quantum efficiency of 25\% and a clock frequency of 125 kHz. In both cases, the after-pulsing probability is less than 8\%.

\vspace{-0.6cm}
\section*{Acknowledgements} 
\vspace{-0.5cm}
The authors would like to thank Chris Michael for his help during the initial stage of this project. This work was supported in part by the startup fund from University of Rochester and by the AFOSR under grant no.~FA9550-12-1-0419. It was performed in part at the Cornell NanoScale Facility, a member of the National Nanotechnology Infrastructure Network, which is supported by the National Science Foundation (Grant ECS-0335765).

\if{
\noindent\textbf{Author Contributions} W.C.J. performed the device modeling and fabrication. X.L. and W.C.J. performed the device testing with support from J.Z. X.L. performed the majority of analysis. J.Z. helped plan the experiment. Q.L. and O.P. conceived the idea and developed the device concept. All authors worked together to write the manuscript.

\noindent\textbf{Author Information} Reprints and permissions are available at www.nature.com/reprints.  The authors declare no competing financial interests.  Correspondence and requests for materials should be sent to Q.L. (qiang.lin@rochester.edu).
}\fi

\end{footnotesize}

\appendix

\vspace{-0.3cm}
\section{Group-velocity dispersion of a silicon microdisk resonator}
\label{AppA}
\vspace{-0.5cm}
For each mode family, the silicon midrodisk resonator is simulated by the finite-element method to find the cavity resonances $\omega(m)$ as a function of mode number $m$ over a broad spectrum. As the propagation constant of a cavity mode is well approximated by $k=m/R$ where R is the device radius, we obtain the dispersion relation $k(\omega)$ for the cavity modes. Fitting it with a high-order polynomial, we obtain the group-velocity dispersion shown in Fig.~1b of the main text.

In general, the group-velocity dispersion of a microdisk resonator is close to a slab waveguide with the same thickness, which can be used as a rough guidance for searching the group-velocity dispersion of a microdisk resonator.

\vspace{-0.3cm}
\section{Theory of photon-pair generation in a microdisk resonator}
\label{AppB}
\begin{figure}[htb]
\includegraphics[width=0.75\columnwidth]{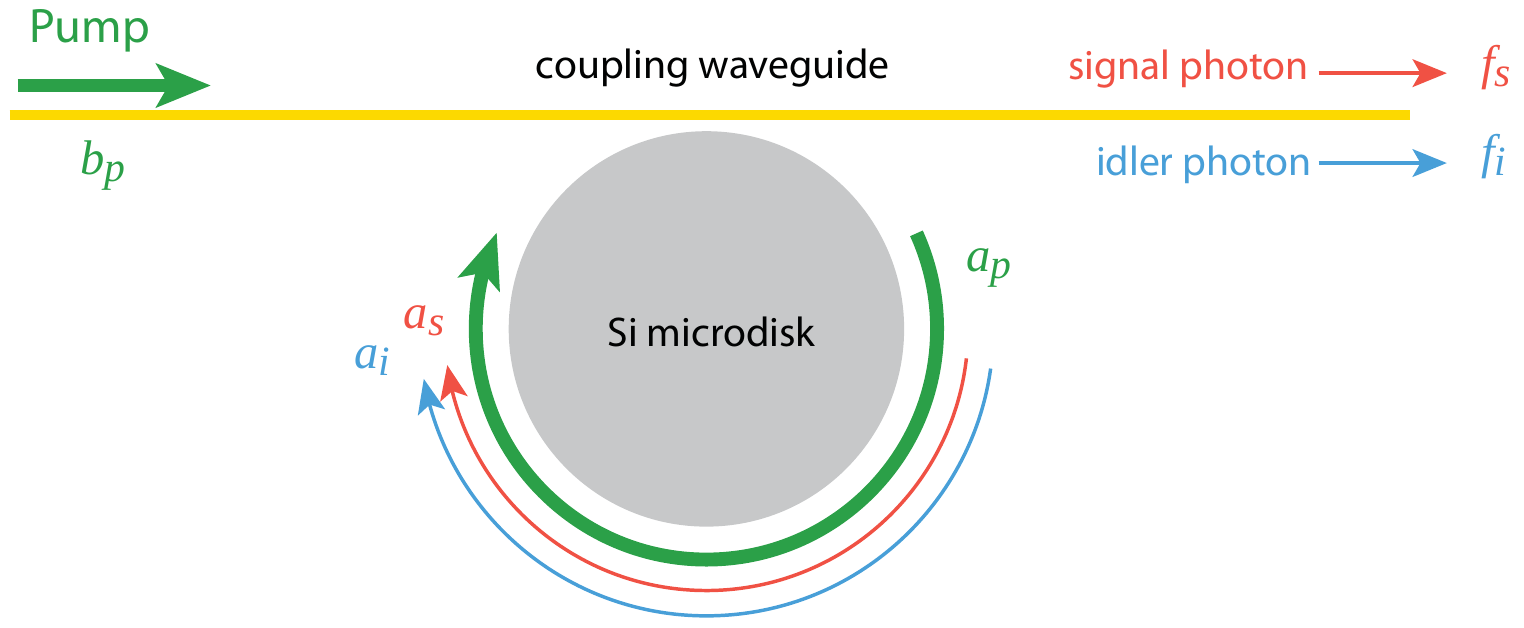}
\caption{\label{FigS1} {\footnotesize Schematic of photon pair generation in a silicon microdisk resonator. }}
\end{figure}

\vspace{-0.5cm}
In this section, we provide the theory describing photon pair generation in a silicon microdisk resonator. We consider only the Kerr nonlinear optical interaction and neglect two-photon absorption and free-carrier effect because of their negligible effect in our device. We consider first the case shown in Fig.~\ref{FigS1} where a pump wave launched into the resonator produces a pair of signal and idler photon which are transmitted out into the coupling waveguide. The Kerr nonlinear interaction among the three cavity modes at frequencies $\omega_{\rm 0j}$ ($j=p,s,i$) can be described by a Hamiltonian $H = H_0 + H_I$, where $H_0$ describes the passive cavity modes coupled to the bus waveguide and $H_I$ describes the Kerr nonlinear interaction \cite{Haus94} inside the cavity, with the following forms
\begin{eqnarray}
H_0 &=& \sum_{j=p,s,i}{\left\{\hbar \omega_{\rm 0j} a_j^\dag a_j - \hbar \sqrt{\Gamma_{\rm ej}} \left[ a_j^\dag b_j e^{-i\omega_j t} + b_j^\dag a_j e^{i\omega_j t} \right]\right\}}, \label{H0} \\
H_I &=& - \frac{\hbar}{2} g_p (a_p^\dag)^2 a_p^2 - 2\hbar a_p^\dag a_p \left( g_{\rm ps} a_s^\dag a_s + g_{\rm pi} a_i^\dag a_i \right) - \hbar g_{\rm psi} a_s^\dag a_i^\dag a_p^2 - \hbar g_{\rm psi}^* (a_p^\dag)^2 a_s a_i, \label{H_I}
\end{eqnarray}
where the intracavity field operator $a_j$ ($j=p,s,i$) is normalized such that $a_j^\dag a_j$ represents the photon number operator and $b_j$ is the field operator of the incoming wave at carrier frequency $\omega_j$ inside the coupling waveguide normalized such that $b_j^\dag b_j$ represents the operator of the input photon flux. $\ave{b_s^\dag b_s}=\ave{b_i^\dag b_i} = 0$ since only the pump is launched into the cavity. $b_j$ satisfies the commutation relation of $[b_j(t), b_j^\dag (t')] = \delta (t-t')$. $\Gamma_{\rm ej}$ is the external coupling rate of the cavity mode at $\omega_{\rm 0j}$.

In Eq.~(\ref{H_I}), the first and second terms describe the self-phase and cross-phase modulation (SPM and XPM) from the pump mode, respectively, and the third and fourth terms govern the four-wave mixing (FWM)  process. We have assumed the signal and idler are much weaker than the pump so that we can neglect the SPM and XPM introduced by them. $g_p$, $g_{\rm pj}$ ($j=s,i$), and $g_{\rm psi}$ are the vacuum coupling rate for the SPM, XPM, and FWM, respectively. They are given by the following general expression
\begin{eqnarray}
g_{\rm ijkl} = \frac{3 \hbar \sqrt{\omega_{\rm 0i} \omega_{\rm 0j} \omega_{\rm 0k} \omega_{\rm 0l}} \eta_{\rm ijkl}}{4 \varepsilon_0 n_i n_j n_k n_l \bar{V}_{\rm ijkl}}  \chi^{(3)}(-\omega_{\rm 0i}; \omega_{\rm 0j}, -\omega_{\rm 0k}, \omega_{\rm 0l}), \label{g}
\end{eqnarray}
where $n_v$ ($v=i,j,k,l$) is the refractive index of silicon material at $\omega_{\rm 0v}$ and $\chi^{(3)}$ is the third-order nonlinear susceptibility. $\bar{V}_{\rm ijkl} = (V_i V_j V_k V_l)^{1/4}$ is the average effective mode volume and $V_v$ ($v=i,j,k,l$) is that at individual frequency $\omega_{\rm 0v}$ given as \cite{Lin08}
\begin{equation}
V_v = \frac{ \left\{ \int d\bm{r} \epsilon_r(\bm{r},\omega_{\rm 0v}) |\widetilde{E}_v(\bm{r},\omega_{\rm 0v})|^2 \right\}^2}{\int_{si} d\bm{r} \epsilon_r^2 (\bm{r},\omega_{\rm 0v}) |\widetilde{E}_v(\bm{r},\omega_{\rm 0v})|^4}, \label{Veff}
\end{equation}
where $\epsilon_r (\bm{r}, \omega_{\rm 0v})$ is the dielectric constant of material and the subscript $si$ in the integral denotes the integral over the silicon core. $\eta_{\rm ijkl}$ describes the spatial mode overlap given by \cite{Lin08}
\begin{equation}
\eta_{\rm ijkl} \equiv \frac{\int_{si} d\bm{r} (\epsilon_{ri} \epsilon_{rj} \epsilon_{rk} \epsilon_{rl})^{1/2} \widetilde{E}_i^* \widetilde{E}_j \widetilde{E}_k^* \widetilde{E}_l}{\left\{ \prod_{v=i,j,k,l} {\int_{si} d\bm{r} \epsilon_{rv}^2 |\widetilde{E}_v|^4} \right\}^{1/4}}, \label{ModeOverlap}
\end{equation}
where $\epsilon_{rv}=\epsilon_r(\bm{r},\omega_{\rm 0v})$ and $\widetilde{E}_v = \widetilde{E}_v(\bm{r},\omega_{\rm 0v})$ is the electrical field profile for the cavity resonant mode at $\omega_{\rm 0v}$.

The vacuum coupling rates for various nonlinear effects are thus obtained from Eq.~(\ref{g}) as $g_p = g_{\rm pppp}$, $g_{\rm pj} = g_{\rm pjjp}$ ($j=s,i$), and $g_{\rm psi} = g_{\rm spip}$. In our device, the three cavity modes are of close frequencies and similar mode profiles, resulting in
\begin{eqnarray}
g_p \approx g_{\rm pj} \approx g_{\rm psi} \equiv g = \frac{c \eta n_2 \hbar \omega_p \sqrt{\omega_s \omega_i}}{n_s n_i \bar{V}}, \label{g_Approx}
\end{eqnarray}
where $\eta$ and $\bar{V}$ denote $\eta_{\rm spip}$ and $\bar{V}_{\rm spip}$, respectively, and we have used the definition of Kerr nonlinear coefficient \cite{Lin07}: $n_2 = \frac{3 \chi^{(3)}}{4\varepsilon_0 c n_p^2}$ at the pump frequency. Using Eqs.~(\ref{H0}) and (\ref{H_I}) and counting in the intrinsic cavity loss \cite{MilburnBook}, we obtain the following equations of motion in the Heisenberg picture governing the wave dynamics inside the cavity:
\begin{eqnarray}
\frac{da_p}{dt} &=& (-i\omega_{\rm 0p} - \Gamma_{\rm tp}/2) a_p + i g a_p^\dag a_p^2 + i \sqrt{\Gamma_{\rm ep}} b_p e^{-i\omega_p t} + i \sqrt{\Gamma_{\rm 0p}} u_p, \label{dapdt} \\
\frac{da_s}{dt} &=& (-i\omega_{\rm 0s} - \Gamma_{\rm ts}/2) a_s + 2 i g a_p^\dag a_p a_s + i g a_i^\dag a_p^2 + i \sqrt{\Gamma_{\rm es}} b_s e^{-i\omega_s t} + i \sqrt{\Gamma_{\rm 0s}} u_s, \label{dasdt} \\
\frac{da_i}{dt} &=& (-i\omega_{\rm 0i} - \Gamma_{\rm ti}/2) a_i + 2 i g a_p^\dag a_p a_i + i g a_s^\dag a_p^2 + i \sqrt{\Gamma_{\rm ei}} b_i e^{-i\omega_i t} + i \sqrt{\Gamma_{\rm 0i}} u_i, \label{daidt}
\end{eqnarray}
where $\Gamma_{\rm 0j}$ and $\Gamma_{\rm tj} = \Gamma_{\rm 0j} + \Gamma_{\rm ej}$ ($j=p,s,i$) are the photon decay rates of the intrinsic and loaded cavity, respectively. $u_j$ is the noise operator associated with intrinsic cavity loss, which satisfies the commutation relation of $[u_j(t), u_j^\dag (t')] = \delta (t-t')$. In Eqs.~(\ref{dapdt})-(\ref{daidt}), we have neglected the pump depletion and the signal/idler-induced SPM and XPM. The transmitted field is given by \cite{MilburnBook}
\begin{eqnarray}
f_j = b_j + i \sqrt{\Gamma_{\rm ej}} a_j. \label{f_j}
\end{eqnarray}

In general, the pump mode can be treated as a classical field and Eqs.~(\ref{dapdt})-(\ref{daidt}) can be easily solved to find the solutions for $a_s$ and $a_i$, from which, together with Eq.~(\ref{f_j}), we find the emitted photon fluxes of signal and idler as
\begin{eqnarray}
\ave{f_s^\dag (t) f_s(t)} &=& \frac{\Gamma_{\rm es} \Gamma_{\rm ti}}{2\pi} \int_{-\infty}^{+\infty}{\left| B(\omega) \right|^2 d\omega}, \label{SignalFlux}\\
\ave{f_i^\dag (t) f_i(t)} &=& \frac{\Gamma_{\rm ei} \Gamma_{\rm ts}}{2\pi} \int_{-\infty}^{+\infty}{\left| B(\omega) \right|^2 d\omega}, \label{IdlerFlux}
\end{eqnarray}
and the pair correlation as
\begin{eqnarray}
p_{c}(t_s,t_i) & \equiv & \ave{f_i^\dag(t_i) f_s^\dag (t_s) f_s(t_s) f_i(t_i)} -\ave{f_s^\dag(t_s) f_s(t_s)} \ave{f_i^\dag(t_i) f_i(t_i)} \nonumber \\
& =& \Gamma_{\rm es} \Gamma_{\rm ei} \left| \frac{1}{2\pi} \int_{-\infty}^{+\infty}{B(-\omega) \left[ \Gamma_{\rm ts} A(\omega) - 1 \right] e^{-i \omega (t_s - t_i)} d\omega } \right|^2, \label{Corr}
\end{eqnarray}
where $A(\omega)$ and $B(\omega)$ have the following expressions
\begin{eqnarray}
A(\omega) &=& \frac{\Gamma_{\rm ti}/2 - i\omega}{(\Gamma_{\rm ts}/2 - i\omega)(\Gamma_{\rm ti}/2 - i\omega) - (g N_p)^2}, \label{alpha}\\
B(\omega) &=& \frac{-i g a_p^2 }{(\Gamma_{\rm ts}/2 - i\omega)(\Gamma_{\rm ti}/2 - i\omega) - (g N_p)^2}, \label{beta}
\end{eqnarray}
where $N_p = \ave{a_p^\dag a_p}$ is the average photon number of the pump wave inside the cavity.

Equations (\ref{SignalFlux})-(\ref{Corr}) are quite general since they include the multiphoton generation induced by the stimulated FWM. For single photon-pair generation, Eqs.~(\ref{SignalFlux}) and (\ref{IdlerFlux}) reduce to
\begin{eqnarray}
\ave{f_s^\dag (t) f_s(t)} & \approx & \frac{2 \Gamma_{\rm es} (g N_p)^2}{\Gamma_{\rm ts} \bar{\Gamma}}, \label{SignalFlux_Approx}\\
\ave{f_i^\dag (t) f_i(t)} & \approx & \frac{2 \Gamma_{\rm ei} (g N_p)^2}{\Gamma_{\rm ti} \bar{\Gamma}}, \label{IdlerFlux_Approx}
\end{eqnarray}
where $\bar{\Gamma} = (\Gamma_{\rm ts} + \Gamma_{\rm ti})/2$ represents the average photon decay rate of the loaded cavity. The photon-pair correlation reduces to
\begin{eqnarray}
  p_{c}(t_s,t_i) \approx \left\{
  \begin{array}{l l}
    \frac{\Gamma_{\rm es} \Gamma_{\rm ei}}{\bar{\Gamma}^2} (g N_p)^2 e^{-\Gamma_{\rm ts} (t_s - t_i)} & \quad \text{if $t_s \ge t_i$}\\
    \frac{\Gamma_{\rm es} \Gamma_{\rm ei}}{\bar{\Gamma}^2} (g N_p)^2 e^{-\Gamma_{\rm ti} (t_i - t_s)} & \quad \text{if $t_s < t_i$}\\
  \end{array} \right. . \label{correlation_Approx}
\end{eqnarray}
Equation (\ref{correlation_Approx}) can be regarded as the joint probability density of emitting a photon pair into the coupling waveguide. It is given as Eq.~(1) in the main text. Integrating Eq.~(\ref{correlation_Approx}) over $\tau = t_s - t_i$, we obtain the average photon-pair emission flux as
\begin{eqnarray}
R_c = \frac{\Gamma_{\rm es} \Gamma_{\rm ei}}{\Gamma_{\rm ts} \Gamma_{\rm ti}} \frac{2 (g N_p)^2}{\bar{\Gamma}}, \label{Rc}
\end{eqnarray}
which is Eq.~(2) in the main text. Comparing Eq.~(\ref{Rc}) with Eqs.~(\ref{SignalFlux_Approx}) and (\ref{IdlerFlux_Approx}), we can find that the photon-pair generation rate inside the cavity is $2(gN_p)^2/\bar{\Gamma}$ and the photon extraction efficiency is $\Gamma_{\rm ej}/\Gamma_{\rm tj}$ ($j=s,i$) for the individual signal and idler mode while it is $\Gamma_{\rm es}\Gamma_{\rm ei}/(\Gamma_{\rm ts} \Gamma_{\rm ti})$ for the correlated photon pair. As $\Gamma_{\rm ej}/\Gamma_{\rm tj}$ is less than 1, the photon-pair flux is smaller than the individual photon fluxes. This is experimentally observed in Fig.~4 in the main text.

\begin{figure}[t!]
\includegraphics[width=0.75\columnwidth]{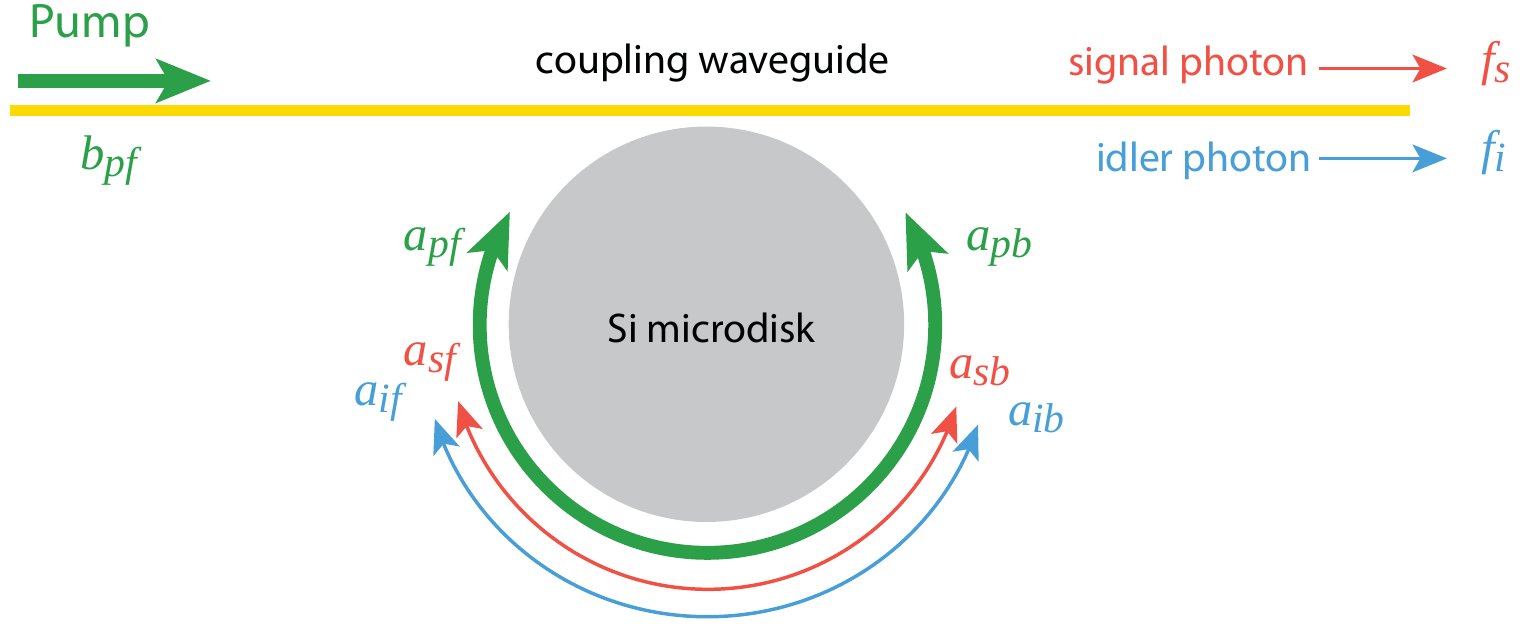}
\caption{\label{FigS2} {\footnotesize Schematic of photon pair generation in a silicon microdisk resonator. The clockwise and counter-clockwise modes are coupled inside the cavity. }}
\end{figure}

In practice, the situation is much more complicated than the ideal situation described above, since, due to the light scattering, the optical wave inside the cavity will couple to the degenerate mode propagating backward in the opposite direction (Fig.~\ref{FigS2}). For a passive cavity, this phenomenon manifests as a doublet in the transmission spectrum (Insets of Fig.~3a in the main text). In this case, The Hamiltonian $H_0$ and $H_I$ become
\begin{eqnarray}
H_0 = && \sum_{j=p,s,i}{ \left\{ \hbar \omega_{\rm 0j} \left(a_{\rm jf}^\dag a_{\rm jf} + a_{\rm jb}^\dag a_{\rm jb} \right) - \left( \hbar \beta_j a_{\rm jf}^\dag a_{\rm jb} + \hbar \beta_j^* a_{\rm jb}^\dag a_{\rm jf} \right) \right.}\nonumber \\
&& {\left. - \hbar \sqrt{\Gamma_{\rm ej}} \left[ (a_{\rm jf}^\dag b_{\rm jf} + a_{\rm jb}^\dag b_{\rm jb}) e^{-i\omega_j t} + (b_{\rm jf}^\dag a_{\rm jf} + b_{\rm jb}^\dag a_{\rm jb} ) e^{i\omega_j t} \right] \right\}}, \label{H0_Doublet} \\
H_I =&& - \frac{\hbar g_p}{2} \left[ (a_{\rm pf}^\dag)^2 a_{\rm pf}^2 + (a_{\rm pb}^\dag)^2 a_{\rm pb}^2 + 4 a_{\rm pf}^\dag a_{\rm pf} a_{\rm pb}^\dag a_{\rm pb} \right] \nonumber\\
 && - 2\hbar (a_{\rm pf}^\dag a_{\rm pf} + a_{\rm pb}^\dag a_{\rm pb} ) \left[ g_{\rm ps} (a_{\rm sf}^\dag a_{\rm sf} + a_{\rm sb}^\dag a_{\rm sb} ) + g_{\rm pi} (a_{\rm if}^\dag a_{\rm if} + a_{\rm ib}^\dag a_{\rm ib} )  \right] \nonumber\\
 && - \hbar g_{\rm psi} \left[ a_{\rm sf}^\dag a_{\rm if}^\dag a_{\rm pf}^2 + a_{\rm sb}^\dag a_{\rm ib}^\dag a_{\rm pb}^2\right] - \hbar g_{\rm psi}^* \left[ (a_{\rm pf}^\dag)^2 a_{\rm sf} a_{\rm if} + (a_{\rm pb}^\dag)^2 a_{\rm sb} a_{\rm ib} \right], \label{H_I_Doublet}
\end{eqnarray}
where $a_{\rm jf}$ and $a_{\rm jb}$ ($j=p,s,i$) represent the optical fields propagating clockwise and counter-clockwise, respectively, inside the cavity (Fig.~\ref{FigS2}), and $\beta_j$ is the coupling coefficient between them. $b_{\rm jf}$ and $b_{\rm jb}$ represent the input fields propagating forward (input from the left end) and backward (input from the right end), respectively, inside the coupling waveguide. The transmitted fields now become $f_j = b_{\rm jf} + i \sqrt{\Gamma_{\rm ej}} a_{\rm jf}$. Using Eqs.~(\ref{H0_Doublet}) and (\ref{H_I_Doublet}) and following the same procedure, we can find the equations of motion as
\begin{eqnarray}
\frac{da_{\rm pf}}{dt} &=& (-i\omega_{\rm 0p} - \Gamma_{\rm tp}/2) a_{\rm pf} + i \beta_p a_{\rm pb} + i g (a_{\rm pf}^\dag a_{\rm pf} + 2 a_{\rm pb}^\dag a_{\rm pb}) a_{\rm pf} + i \zeta_{\rm pf}, \label{dapfdt_Doublet} \\
\frac{da_{\rm pb}}{dt} &=& (-i\omega_{\rm 0p} - \Gamma_{\rm tp}/2) a_{\rm pb} + i \beta_p^* a_{\rm pf} + i g (a_{\rm pb}^\dag a_{\rm pb} + 2 a_{\rm pf}^\dag a_{\rm pf}) a_{\rm pb} + i \zeta_{\rm pb}, \label{dapbdt_Doublet} \\
\frac{da_{\rm sf}}{dt} &=& (-i\omega_{\rm 0s} - \Gamma_{\rm ts}/2) a_{\rm sf} + i \beta_s a_{\rm sb} + 2 i g (a_{\rm pf}^\dag a_{\rm pf} + a_{\rm pb}^\dag a_{\rm pb} ) a_{\rm sf} + i g a_{\rm if}^\dag a_{\rm pf}^2 + i \zeta_{\rm sf}, \label{dasfdt_Doublet} \\
\frac{da_{\rm sb}}{dt} &=& (-i\omega_{\rm 0s} - \Gamma_{\rm ts}/2) a_{\rm sb} + i \beta_s^* a_{\rm sf} + 2 i g (a_{\rm pf}^\dag a_{\rm pf} + a_{\rm pb}^\dag a_{\rm pb} ) a_{\rm sb} + i g a_{\rm ib}^\dag a_{\rm pb}^2 + i \zeta_{\rm sb}, \label{dasbdt_Doublet} \\
\frac{da_{\rm if}}{dt} &=& (-i\omega_{\rm 0i} - \Gamma_{\rm ti}/2) a_{\rm if} + i \beta_i a_{\rm ib} + 2 i g (a_{\rm pf}^\dag a_{\rm pf} + a_{\rm pb}^\dag a_{\rm pb} ) a_{\rm if} + i g a_{\rm sf}^\dag a_{\rm pf}^2 + i \zeta_{\rm if}, \label{daifdt_Doublet} \\
\frac{da_{\rm ib}}{dt} &=& (-i\omega_{\rm 0i} - \Gamma_{\rm ti}/2) a_{\rm ib} + i \beta_i^* a_{\rm if} + 2 i g (a_{\rm pf}^\dag a_{\rm pf} + a_{\rm pb}^\dag a_{\rm pb} ) a_{\rm ib} + i g a_{\rm sb}^\dag a_{\rm pb}^2 + i \zeta_{\rm ib}, \label{daibdt_Doublet}
\end{eqnarray}
where $\zeta_{\rm jv} \equiv \sqrt{\Gamma_{\rm ej}} b_{\rm jv} e^{-i\omega_j t} + \sqrt{\Gamma_{\rm 0j}} u_{\rm jv}$ ($j=p,s,i$ and $v=f,b$).

Similar to the previous case, Eqs.~(\ref{dapfdt_Doublet})-(\ref{daibdt_Doublet}) can be used to find the solution of signal and idler fields inside the cavity $a_{\rm jv}$ ($j=s,i$ and $v=f,b$), through which we obtain the signal and idler fluxes as
\begin{eqnarray}
\ave{f_s^\dag (t) f_s(t)} &=& \frac{\Gamma_{\rm es} \Gamma_{\rm ti}}{2\pi} \int_{-\infty}^{+\infty}{\left[ \left| M_{\rm 13}(\omega) \right|^2 + \left| M_{\rm 14}(\omega) \right|^2 \right] d\omega}, \label{SignalFlux_Doublet}\\
\ave{f_i^\dag (t) f_i(t)} &=& \frac{\Gamma_{\rm ei} \Gamma_{\rm ts}}{2\pi} \int_{-\infty}^{+\infty}{\left[ \left| M_{\rm 31}(\omega) \right|^2 + \left| M_{\rm 32}(\omega) \right|^2 \right] d\omega}, \label{IdlerFlux_Doublet}
\end{eqnarray}
and the pair correlation as
\begin{eqnarray}
&& p_{c}(t_s,t_i) \equiv \ave{f_i^\dag(t_i) f_s^\dag (t_s) f_s(t_s) f_i(t_i)} -\ave{f_s^\dag(t_s) f_s(t_s)} \ave{f_i^\dag(t_i) f_i(t_i)} \nonumber \\
&& = \Gamma_{\rm es} \Gamma_{\rm ei} \left| \frac{1}{2\pi} \int_{-\infty}^{+\infty}{ \left\{\Gamma_{\rm ts} \left[ M_{\rm 11}(\omega) M_{\rm 31}^*(\omega) + M_{\rm 12}(\omega) M_{\rm 32}^*(\omega) \right] - M_{\rm 31}^*(\omega) \right\} e^{-i \omega (t_s - t_i)} d\omega } \right|^2, \qquad \label{Corr_Doublet}
\end{eqnarray}
where $M_{\rm jk}$ is the elements of the matrix $M(\omega)$ which is given by
\begin{equation}
 M^{-1}(\omega) = -
 \begin{pmatrix}
  i\omega - \Gamma_{\rm ts}/2 & i \beta_s & -i g a_{\rm pf}^2 & 0 \\
  i \beta_s^* & i\omega - \Gamma_{\rm ts}/2 & 0 & -i g a_{\rm pb}^2 \\
  i g^* (a_{\rm pf}^*)^2  & 0  & i\omega - \Gamma_{\rm ti}/2 & - i \beta_i^*  \\
  0 & i g^* (a_{\rm ab}^*)^2 & - i \beta_i & i\omega - \Gamma_{\rm ti}/2
 \end{pmatrix}. \label{M_Matrix}
\end{equation}

Equations (\ref{SignalFlux_Doublet}) and (\ref{IdlerFlux_Doublet}) provide the theoretical curves of individual signal and idler fluxes given in Fig.~4a-d in the main text. Equation (\ref{Corr_Doublet}) shows that the photon-pair correlation is only a function of time difference $\tau = t_s - t_i$, $p_{c}(t_s,t_i) = p_{c}(\tau)$. Integrated $p_{c}(\tau)$ over $\tau$, we obtain the photon-pair emission flux $R_c = \int_{-\infty}^{+\infty}{p_{c}(\tau)d\tau}$, which is given as the theoretical curves in Fig.~4e-h in the main text. The complete expressions of Eq.~(\ref{g})-(\ref{ModeOverlap}) are used in the theoretical calculation, where the effective mode volumes and the mode overlap factors are obtained from the finite element simulation of the device. A Kerr nonlinear coefficient, $n_2 = 5.0 \times 10^{-5}~{\rm cm^2/GW}$, is used in the theoretical calculation. The cavity parameters, such as $\Gamma_{\rm 0j}$, $\Gamma_{\rm ej}$, and $\beta_j$ ($j=p,s,i$), are measured from the laser-scanned transmission spectrum of the passive cavity. The ${\rm P_1}$ mode at 1584.1~nm is outside the scanning range of our lasers and its cavity parameters cannot be obtained through independent measurement. We approximate its intrinsic Q and mode splitting with those of the nearest ${\rm P_1}$ mode at 1564.9~nm since the properties of cavity modes remain similar within each mode family. Moreover, we approximate its external coupling with that of the ${\rm P_2}$ mode at 1553.6~nm because of their similar external couplings as observed from the PL spectra shown in Fig.~3b and c of the main text. Figure 4d,h,i of the main text show that these parameters provide a good approximation for the cavity properties of the ${\rm P_1}$ mode at 1584.1~nm.

\vspace{-0.3cm}
\section{Statistics of coincidence photon counting}
\label{AppC}
\begin{figure}[t!]
\includegraphics[width=0.8\columnwidth]{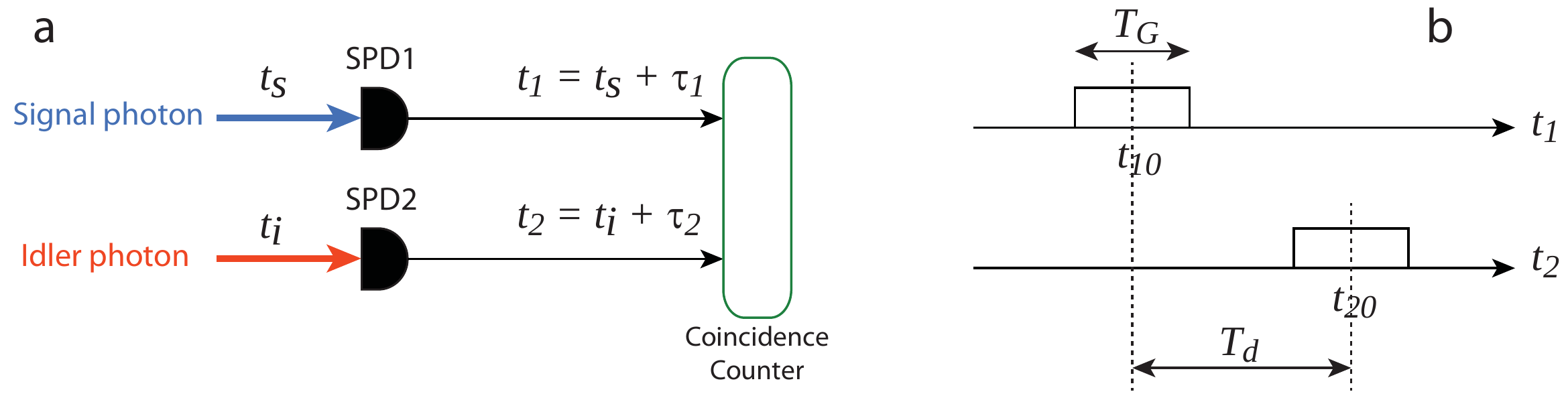}
\caption{\label{FigS3} {\footnotesize {\bf a.} Schematic of coincidence photon counting. A signal and idler photon arrive at the detectors at time $t_s$ and $t_i$, respectively, with a probability density of $p_{ph}(t_s,t_i)$. The two detectors produce a pair of events which are recorded by the coincidence counter at time $t_1$ and $t_2$, with a probability density of $p(t_1,t_2)$. {\bf b.} The time relationship of the events arriving at the coincidence counter. Each event arrives within a gating window centered at $t_{j0}$ ($j=1,2$) and with a width of $T_G$. The two input channels of the coincidence counter has a time delay of $T_d$. }}
\end{figure}

\vspace{-0.5cm}
In this section, we provide the theoretical description of the statistics of coincidence photon counting with gated single photon detectors (SPDs). Assume a signal and idler photon strike the detectors at time $t_s$ and $t_i$ with a probability density of $p_{ph}(t_s, t_i)$. The SPDs then produce a pair of photoelectric events which are recorded by the coincidence counter at a later time $t_1 = t_s + \tau_1$ and $t_2 = t_i + \tau_2$ (Fig.~\ref{FigS3}), where $\tau_1$ and $\tau_2$ include the electron transit time inside the SPDs, the propagation time from the SPDs to the coincidence counter, and time delays introduced by the coincidence counter, with a joint probability density of $p_J(\tau_1, \tau_2)$. The joint probability density of the photoelectric events is thus given by
\begin{eqnarray}
p_{SI}(t_1,t_2) &=& \xi_1 \xi_2 \dint_{-\infty}^{+\infty}{p_{ph}(t_s,t_i) p_J(\tau_1,\tau_2) d\tau_1 d\tau_2} \nonumber \\
&=& \xi_1 \xi_2 \int_{-\infty}^{+\infty}{p_{ph}(t_2-t_1-\tau_D) p_{\rm IRF}(\tau_D) d\tau_D}, \label{p_SI}
\end{eqnarray}
where $\xi_j$ ($j=1,2$) are the detection efficiencies of the two SPDs, and we have used the fact that $p_{ph}(t_s, t_i) = p_{ph}(t_i-t_s)$ which is only a function of the time difference between the signal and idler arrival. For a correlated photon pair, Eq.~(\ref{Corr_Doublet}) indicates that $p_{ph}(t_s,t_i) = \alpha_s \alpha_i \ave{f_i^\dag(t_i) f_s^\dag (t_s) f_s(t_s) f_i(t_i)}$ where $\alpha_s$ and $\alpha_i$ are the propagation losses of signal and idler photons from the coupling waveguide (a tapered fiber in our case) to the SPDs. In Eq.~(\ref{p_SI}), $p_{\rm IRF}(\tau)$ is the instrument response function (IRF) of the coincidence counting system, given as
\begin{eqnarray}
p_{\rm IRF}(\tau_D) \equiv \int_0^{+\infty} {p_J(\tau_1,\tau_1 + \tau_D) d\tau_1}. \label{p_IRF}
\end{eqnarray}
The IRF is independently measured by launching identical ultrashort pulses onto the SPDs and recording the coincidence spectrum (see Method in the main text). In out coincidence counting system, the IRF is dominantly induced by the timing jitters of the two SPDs. As the detector timing jitter varies with its quantum efficiency (QE), the IRF is recorded at different QEs for specific experiments (QE is 15\% and 25\% for Fig.~4 and 5 in the main text, respectively). In general, the experimentally recorded IRF is well described by a Gaussian function since the timing jitters of the SPDs follow a Gaussian distribution.

As a detection event can be either a photoelectric event fired by a photon or a detector dark count, the joint probability density, $p(t_1,t_2)$, that the two detectors produce a pair of events at time $t_1$ and $t_2$ (Fig.~\ref{FigS3}) is thus given by
\begin{eqnarray}
p(t_1,t_2) = p_{SI}(t_1,t_2) + p_{SD}(t_1,t_2) + p_{DI}(t_1,t_2) + p_{DD}(t_1,t_2), \label{p_Det}
\end{eqnarray}
where $p_{SD}$, $p_{DI}$, and $p_{DD}$ are the joint probability densities of an event pair of signal$-$dark count, dark count$-$idler, and dark count$-$dark count, respectively, produced in the two separate SPDs. The generation of a detector dark count is independent of an event generated in another detector (either a photoelectric event produced by an incoming photon or a detector dark count). As a result, $p_{SD}(t_1,t_2) = p_{S}(t_1) p_{D2}(t_2)$, $p_{DI}(t_1,t_2) = p_{D1}(t_1) p_{I}(t_2)$, and $p_{DD}(t_1,t_2) = p_{D1}(t_1) p_{D2}(t_2)$, where $p_S(t)$, $p_I(t)$, and $p_{Dj}(t)$ ($j=1,2$) are the probability densities of detection events from signal, idler, and dark counts, respectively, produced in the two detectors. In experiment, $p_S(t)$, $p_I(t)$, and $p_{Dj}(t)$ are equal to the average detection rates of the related events, which can be measured independently. Consequently, Eq.~(\ref{p_Det}) becomes
\begin{eqnarray}
p(t_1,t_2) = p_{SI}(t_2-t_1) + p_{S}(t_1)p_{D2}(t_2) + p_{D1}(t_1)p_I(t_2) + p_{D1}(t_1)p_{D2}(t_2). \label{p_Det2}
\end{eqnarray}
Therefore, the coincidence at $\tau = t_2 - t_1$ within a small delay-time interval of $d\tau$ is given by
\begin{eqnarray}
\rho (\tau) d\tau &=& \int_{t_{10}-T_G/2}^{t_{10}+T_G/2}{dt_1 \int_{t_{20}-T_G/2}^{t_{20}+T_G/2} {dt_2 p(t_1, t_2)}} \nonumber\\
&=& d\tau  \left(T_G - |\tau - T_d| \right) \left[ p_{SI}(\tau - T_d) + p_S p_{D2} + p_{D1} p_I + p_{D1}p_{D2} \right], \label{rho}
\end{eqnarray}
for $|\tau - T_d| \le T_G$ and $\rho(\tau)=0$ otherwise. In experiment, $d\tau$ corresponds to the time resolution of the coincidence counter, which is 4~ps in our system. $t_{10}$ and $t_{20}$ are the centers of the gating windows in the two channels (Fig.~\ref{FigS3}). As a result, the delay-time-dependent coincidence counting histogram (we term as the coincidence spectrum here for simplicity) is obtained accordingly as
\begin{equation}
C(\tau) d\tau = R_G T \rho(\tau) d\tau, \label{C}
\end{equation}
where $R_G$ is the clock frequency of the SPDs and $T$ is the data collection time.

Equations (\ref{p_SI}), (\ref{rho}), and (\ref{C}) provide a complete description of coincidence photon counting, from which we can obtain the true coincidence spectrum $C_T(\tau)$ and the accidence coincidence spectrum $C_A(\tau)$ as
\begin{eqnarray}
C_T(\tau) d\tau &=& R_G T d\tau  \left(T_G - |\tau - T_d| \right) \left[p_{SI}(\tau - T_d) - p_S p_I \right] \nonumber \\
&=& R_G T d\tau  \left(T_G - |\tau - T_d| \right) \xi_1 \xi_2 \alpha_s \alpha_i \int_{-\infty}^{+\infty}{p_{c}(\tau - T_d -\tau_D) p_{\rm IRF}(\tau_D) d\tau_D} , \label{C_T}\\
C_A(\tau) d\tau &=& R_G T d\tau  \left(T_G - |\tau - T_d| \right) (p_S + p_{D1}) (p_I + p_{D2}), \label{C_A}
\end{eqnarray}
where $p_c(\tau)$ is given in Eq.~(\ref{Corr_Doublet}). The theoretical coincidence-to-accidental ratio (CAR) shown as the red curves in Fig.~4e-h of the main text is obtained as
\begin{eqnarray}
{\rm CAR} = \frac{\int_{- T_{\rm FWHM}/2}^{+ T_{\rm FWHM}/2}{C_T(\tau + T_d) d\tau}}{\int_{- T_{\rm FWHM}/2}^{+ T_{\rm FWHM}/2}{C_A(\tau + T_d) d\tau}}, \label{CAR}
\end{eqnarray}
where $T_{\rm FWHM}$ is the full width at half maximum of the coincidence spectrum. In Eqs.~(\ref{C_T}) and (\ref{C_A}), $p_S = \xi_1 \alpha_s \ave{f_s^\dag (t) f_s(t)} + p_{R_s}$ and $p_I = \xi_2 \alpha_i \ave{f_i^\dag (t) f_i(t)} + p_{R_i}$, where $\ave{f_s^\dag (t) f_s(t)}$ and $\ave{f_i^\dag (t) f_i(t)}$ are given by Eqs.~(\ref{SignalFlux_Doublet}) and (\ref{IdlerFlux_Doublet}), and $p_{R_s}$ and $p_{R_i}$ are the probability densities of detection events from Raman noise photons at the signal and idler channels, respectively (See the next section for the details of characterizing Raman noises). On the other hand, Eq.~(\ref{C_A}) shows clearly that the accidental coincidence can be accurately calculated from independently measured $p_S$, $p_I$, and $p_{Dj}$ ($j=1,2$). Therefore, using these measured parameters together with independently recorded true coincidence we can reproduce the experimentally recorded CAR. This theoretical approach explains well the saturation of CAR at low pair fluxes and is commonly used in literature \cite{Takesue10}. It is shown as the black dashed curves in Fig.~4e-h of the main text.

The theoretical curves shown in Fig.~5 of the main text are described by Eq.~(\ref{C}) (normalized by its peak value), in which $p_{ph}(t_s,t_i) = \alpha_s \alpha_i \ave{f_i^\dag(t_i) f_s^\dag (t_s) f_s(t_s) f_i(t_i)}$ obtained from Eq.~(\ref{Corr_Doublet}) and $p_{\rm IRF}(\tau)$ is given by a Gaussian fitting to the experimentally recorded IRF. The cavity parameters used in $\ave{f_i^\dag(t_i) f_s^\dag (t_s) f_s(t_s) f_i(t_i)}$, such as $\Gamma_{\rm 0j}$, $\Gamma_{\rm ej}$, and $\beta_j$ ($j=p,s,i$), were independently measured from the laser-scanned transmission spectrum of the passive cavity.

\vspace{-0.3cm}
\section{Characterization of Raman noise produced in delivery fibers}
\label{AppD}
\vspace{-0.5cm}
\begin{figure}[hbt]
\includegraphics[width=1.0\columnwidth]{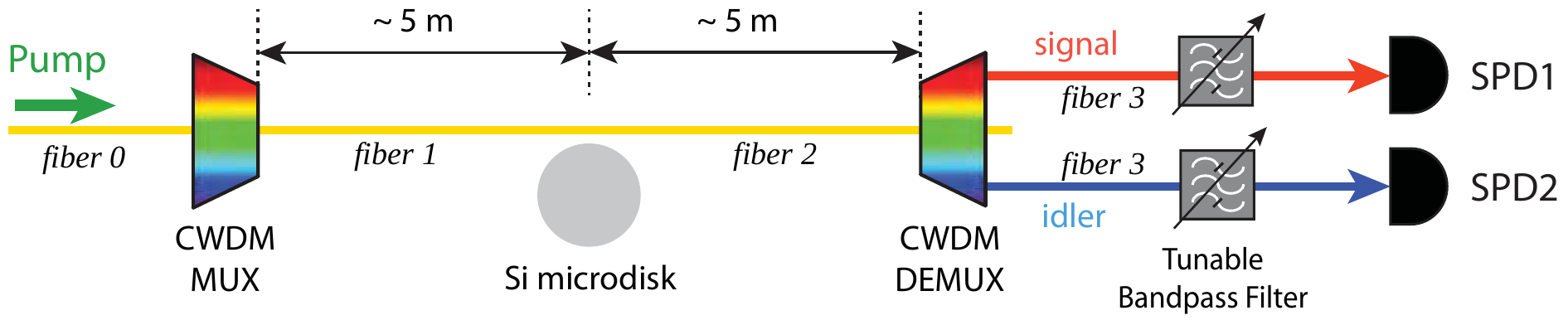}
\caption{\label{FigS5} {\footnotesize Schematic of the experimental setup. Details can be found in Fig.~2 of the main text. }}
\end{figure}
To characterize the noise for each photon mode, we recorded the photon flux when the center wavelength of the tunable bandpass filter was tuned away from the photon mode so that the latter was completely blocked (see Fig.~\ref{FigS5}). The pumping condition remained the same as what we used to characterize the photon pairs, with the tapered fiber touched down to the nanoforks and the pump wavelength on the cavity resonance. Such an experimental condition ensures that the pump power distribution inside the delivery fibers remains unchanged. The photon flux recorded under this condition is not produced from the device since the photon wavelength does not coincide with any cavity resonance.

\begin{figure}[htb]
\includegraphics[width=0.6\columnwidth]{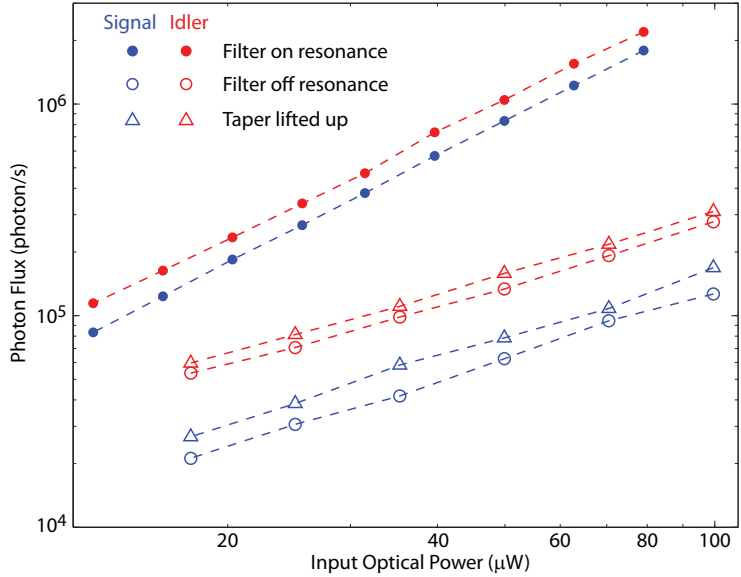}
\caption{\label{FigS4} {\footnotesize Photon fluxes recorded for ${\rm P_2^I}$, with detector dark counts subtracted. The solid and open circles show the photon fluxes when the center wavelengths of the bandpass filters are tuned onto and away from the ${\rm P_2^I}$ mode pair, respectively, with the tapered fiber touched down to the nanoforks. The open triangles are recorded when the tapered fiber is lifted up and thus does not couple to the device. The blue and red show the signal at shorter wavelength and idler at longer wavelength, respectively. The dashed lines are only for eye guidance.}}
\end{figure}

Figure~\ref{FigS4} shows an example of recorded photon fluxes for ${\rm P_2^I}$, where the solid and open circles show the cases when the filter centers were tuned onto and away from the photon mode pair, respectively. The noise photon fluxes exhibit a linear dependence on the pump power. Figure \ref{FigS4} shows clearly that the observed noise is the dominant noise source accompanied with the photon pairs. To further verify that the observed noise does not come from the device, we directly lifted the taper up so that the tapered fiber did not couple to the device and the pump wave propagated only inside the delivery fibers (as well as the optical components in the experimental system. See Fig.~\ref{FigS5}). Kept other conditions unchanged and recorded again the photon fluxes, we obtained the open triangles shown in Fig.~\ref{FigS4}, which are very similar to the open circles. This clearly verifies that the noise photons are not generated from the device but produced inside the experimental testing system.

In our testing system (Fig.~\ref{FigS5}), the CWDM MUX has a band isolation greater than 120 dB. As a result, any broadband noise produced before the MUX (\emph{e.g.}, in \emph{fiber 0}) will be completely blocked. On the other hand, the CWDM DEMUX separates the pump wave from the photon pairs and no noise will be produced in \emph{fiber 3} after the DEMUX. Moreover, when the pump wave is nearly critically coupled into the device, the pump power in \emph{fiber 2} after the device is small. Therefore, the noise should be dominantly produced inside \emph{fiber 1} before the device, which has a rough length of $\sim$5~m. The open triangles in Fig.~\ref{FigS4} have higher magnitudes than the open circles simply because, when the taper is lifted up, the pump power will increase in \emph{fiber 2} and will start to produce noise.

\begin{figure}[b!]
\includegraphics[width=0.75\columnwidth]{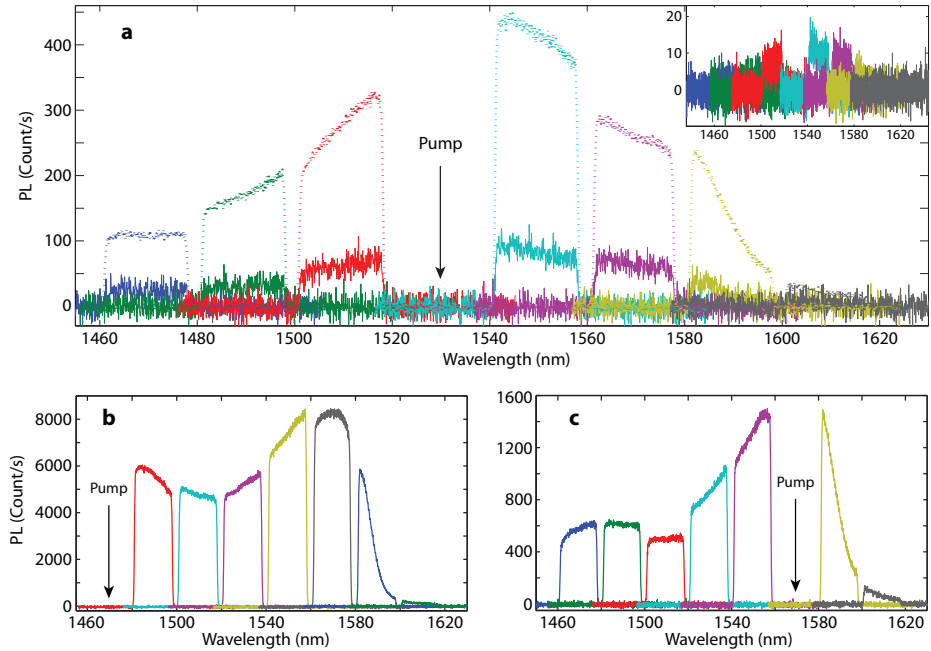}
\caption{\label{FigS6} {\footnotesize {\bf a.} Raman noise spectra recorded with different lengths of \emph{fiber 1}, with an input pump power of about 270~${\rm \mu W}$. The spectra were recorded at different transmission ports of the CWDM DEMUX (indicated by color), for easy suppression of the pump. The solid curves and dotted curves show the cases with a fiber length of $\sim$29~m and $\sim$1~km, respectively. The amplitudes of the dotted curves were reduced by a factor of 10, for easy comparison with the thin solid curves. The inset shows the case with a fiber length of $\sim$5~m, same as what we used to characterize correlated photon pairs. The pump power and wavelength are the same for these three cases. The spectral cut-off beyond $\sim$1590nm is due to the quantum-efficiency cut-off of the detector used in our sepctrometer. {\bf b.} Raman noise spectra with the pump located at 1470nm, showing the Stokes side. \emph{fiber 1} has a length of $\sim$1~km. {\bf c.} Raman noise spectra with the pump located at 1570nm, showing mostly the anti-Stokes side. \emph{Fiber 1} has a length of $\sim$1~km. }}
\end{figure}

Figure \ref{FigS4} shows that the idler channel at the Stokes side of the pump has a larger noise photon flux than the signal channel at the anti-Stokes side. This feature together with their linear power dependence implies that the noise is likely to be the Raman noise produced inside \emph{fiber 1}. To verify this, we incremented the length of \emph{fiber 1} (with the tapered fiber not coupled to the device) and monitored the noise spectrum. Figure \ref{FigS6}a shows an example of the noise spectra recorded with three different fiber lengths of $\sim$5~m, $\sim$29~m, and $\sim$1~km. The similarity of the spectra among these three cases indicates clearly that they share an exactly same physical origin. The noise spectra show the Raman noise characteristics of silica \cite{Tick97}, with the Boson peak clearly visible at a frequency shift of $\sim$1.5~THz away from the pump.

To reveal more characteristics of the noise spectra, we tuned the pump wavelength to the left end of the CWDM and recorded again the noise spectra (Fig.~\ref{FigS6}b). In this case, the noise spectra primarily show the Stokes side. Both the Boson peak and Raman gain peak are clearly visible, with a frequency shift of about 1.5 and 13 THz, respectively, away from the pump frequency. Tuning the pump wavelength to 1570nm we recorded the noise spectra now mostly on the anti-Stokes side (Fig.~\ref{FigS6}c). Again, we observed both peaks but with smaller amplitudes. These observations clearly verify that the observed noise indeed comes from the spontaneous Raman scattering inside the delivery fiber.

As the Raman noise is very broadband, in experiment, we used a pair of tunable bandpass filters with a 3-dB bandwidth of 1.2 nm in front of the SPDs (Fig.~\ref{FigS5}) to cut the Raman noise. The Raman noise fluxes associated with each mode pair were measured with the filter centers tuned away from the mode pair far enough that the latter were completely blocked. The recorded Raman noise fluxes were used in Eq.~(\ref{C_A}) for finding the theoretical CAR, as discussed in the previous section. Due to the strong frequency dependence of Raman noise spectrum, a certain uncertainty may be accompanied with the recorded Raman noise fluxes. Note that the Raman noise can be further suppressed in the future by using narrower filters since the correlated photon pairs have much narrower spectra. The noise can also be reduced by shortening the length of \emph{fiber 1}. Such Raman noise will be completely absent if an on-chip laser \cite{Bowers10} is used as the pump.

\end{document}